\documentclass[reprint,aps,longbibliography,nofootinbib,floatfix,nobalancelastpage]{revtex4-2}
\usepackage{amsmath}
\usepackage{amsfonts}
\usepackage{amssymb}
\usepackage{mathtools}
\usepackage{graphicx}
\usepackage[dvipsnames]{xcolor}
\usepackage[colorlinks=True,citecolor=gray,linkcolor=myRed,urlcolor=gray,hypertexnames=false]{hyperref}
\usepackage{hypcap}
\usepackage{yfonts}
\usepackage{bm}
\usepackage[normalem]{ulem}
\usepackage{dsfont}
\usepackage{braket}
\usepackage{marginnote}
\usepackage{adjustbox}
\usepackage{tikz}
\usepackage{quantikz}
\usetikzlibrary{shapes, arrows.meta}
\newcommand\cbox[1]{\vcenter{\hbox{#1}}}
\usepackage{kantlipsum}

\newcommand{\nocontentsline}[3]{}
\newcommand\stoptoc{%
   \let\origcontentsline\addcontentsline
   \let\addcontentsline\nocontentsline
}
\newcommand\resumetoc{%
   \let\addcontentsline\origcontentsline
}

\newcommand\eq[1]{\begin{align}#1\end{align}}
\newcommand{\eref}{{\cal E}_{\rm ref}}

\newcommand\new[1]{{\color{black}{#1}}}
\newcommand\news[1]{{\color{black}{#1}}}

\newcommand\mytitle{Do mixed states exhibit deep thermalisation?}

\definecolor{myBlue}{RGB}{31,119,180}
\definecolor{myOrange}{RGB}{255,127,14}
\definecolor{myGreen}{RGB}{44,160,44}
\definecolor{myRed}{RGB}{214,39,40}
\definecolor{myPurple}{RGB}{148,103,189}

\begin{document}

\title{\mytitle}

\author{Alan Sherry}
\email{alan.sherry@icts.res.in}
\affiliation{International Centre for Theoretical Sciences, Tata Institute of Fundamental Research, Bengaluru 560089, India}

\author{Sthitadhi Roy}
\email{sthitadhi.roy@icts.res.in}
\affiliation{International Centre for Theoretical Sciences, Tata Institute of Fundamental Research, Bengaluru 560089, India}

\begin{abstract}
Deep thermalisation -- where ensembles of pure states on a local subsystem, conditioned on measurement outcomes on its complement, approach universal maximum-entropy ensembles -- represents a stronger form of ergodicity than conventional thermalisation. We show that this framework fails dramatically for mixed initial states, evolved unitarily, even with infinitesimal initial mixedness. To address this, we introduce a new paradigm of deep thermalisation for mixed states, fundamentally distinct from that for pure-state ensembles. In our formulation, the deep thermal ensemble arises by tracing out auxiliary degrees of freedom from a maximum-entropy ensemble defined on an augmented system, with the ensemble structure depending explicitly on the entropy of the initial state. We demonstrate that such ensembles emerge dynamically in generic, locally interacting chaotic systems. For the self-dual kicked Ising chain, which we show to be exactly solvable for a class of mixed initial states, we find exact emergence of the so-defined mixed-state deep thermal ensemble at finite times. Our results therefore lead to fundamental insights into how maximum entropy principles and deep thermalisation manifest themselves in unitary dynamics of states with finite entropy.
\end{abstract}

\maketitle

Understanding the out-of-equilibrium dynamics of complex quantum systems remains a central question in modern physics, bridging quantum many-body theory and quantum information science.
This question has gained renewed significance with the advent of quantum computing platforms, albeit noisy, that operate far from equilibrium~\cite{preskill2018quantum,boixo2018characterising,smith2019simulating,ippoliti2021many,hoke2023measurement,fauseweh2024quantum}.
The answers to this question not only underlie the emergence of universal thermal ensembles under dynamics~\cite{deutsch1991quantum,*deutsch2018eigenstate,srednicki1994chaos,rigol2008thermalisation,dalessio2016from}
but also shed light onto pertinent questions in quantum information science, such as those related to the emergence of quantum designs~\cite{divincenzo2002quantum,gross2007evenly,roberts2017chaos} and the dynamics of quantum information in general~\cite{lewis2019dynamics}.

Conventional thermalisation is a statement about the emergence of thermal density matrices describing the state of local subsystems, with the rest of the system acting as a bath in the context of isolated quantum systems undergoing unitary dynamics~\cite{deutsch1991quantum,*deutsch2018eigenstate,srednicki1994chaos,rigol2008thermalisation,dalessio2016from}. Formally, the latter is equivalent to tracing out the degrees of freedom in the bath.
However, modern experiments allow for microscopic interrogation of the systems by enabling measurements of non-local, multipartite correlations as well as keeping a record of the measurement outcome~\cite{bakr2009gasmicroscope,haller2015single,neill2016ergodic,kaufman2016quantum,bernien2017simulator,ebadi2021quantum,opremcak2021high,evered2023high,moses2023trappedions,shaw2024universal,choi2023preparing}.
Such measurements, when restricted to the bath, induce an ensemble of states on the local subsystem conditioned on the outcomes of the measurement on the bath. 
The ensemble, dubbed the {\it projected ensemble} (PE)~\cite{choi2023preparing,cotler2023emergent}, manifestly contains a lot more information than the density matrix of the subsystem as the former describes an entire distribution of states whereas the latter is just the first moment of the said distribution.

An important fallout is the extension of the ideas of thermalisation to that of {\it deep thermalisation}~\cite{ho2022exact,ippoliti2022solvablemodelofdeep,cotler2023emergent,choi2023preparing,lucas2023freefermiondeepthm,ippolitu2023dynamical,bhore2023deepthmconstrained,varikuti2024unravelingemergence,chan2024pe,mark2024maximum,manna2025peconserved} which is the phenomenon that the PE maximises the ensemble entropy~\cite{mark2024maximum} which quantifies how well does the ensemble explore the available Hilbert space.
In the absence of any conservation laws, the relevant ensemble is the Haar ensemble which leads to the emergence of quantum designs~\cite{choi2023preparing,cotler2023emergent}. Conservation laws modify the deep thermal ensemble by distorting the Haar ensemble to generate the so-called Scrooge ensemble which is the maximum entropy ensemble of pure states given a density matrix as the first moment of the ensemble~\cite{jozsa1994accessinfo,goldstein2006distribution,lucas2023freefermiondeepthm,chang2024gaussian,chang2025charge,manna2025peconserved,bejan2025matchgate}.

The plethora of recent work on these questions have focussed almost exclusively on the pure states undergoing unitary dynamics. 
In this Letter, we investigate the hitherto unexplored territory of the fate of deep thermalisation when the initial state itself is mixed, or in other words, has finite entropy.
This question is of fundamental interest as the nature of the induced PEs can be completely different from those in the case of pure states; in particular, the maximum entropy principle itself for ensembles of mixed states is not clearly established~\cite{milekhin2024observableprojectedensembles,sherry2025miqc,yu2025mixedstatedeepthermalization}.
At the same time, the question is of increasingly practical importance as the presence of finite entropy due to state preparation and measurement (SPAM) errors is ubiquitous in all experimental platforms~\cite{moses2023trappedions,bernien2017simulator}.

In this Letter we show that deep thermalisation as defined for pure states is fragile to any amount of mixedness in the initial state even for thermodynamically large baths and maximally chaotic dynamics. 
Instead the asymptotic projected ensemble of mixed states is described by partially tracing out pure states from a maximum entropy ensemble, which depends on the initial state, over a system augmented by auxiliary degrees of freedom and hence defines a new paradigm for deep thermalisation of mixed states; this is shown by analytically computing arbitrary moments of the said ensembles and constitutes the first result of this work.
The second key result is the demonstration of the emergence of this ensemble for locally interacting chaotic dynamics. In particular, we show analytically for an exactly solvable model that the mixed-state ensemble emerges at a finite time. We complement our exact results with numerical results away from the exactly solvable point.

To describe and explain these results concretely, let us lay out some definitions explicitly. 
Consider a system bipartitioned into a subsystem $A$ and its complement, $B$. The PE will be induced on the former due to measurements on the latter.
Given a generic state $\varrho$ of the system, the PE,  denoted by 
\eq{
{\cal E}_{\rm PE}[\varrho] :=\{p(z_B), \varrho_A(z_B)\}_{z_B}\,,
\label{eq:PE-def}
}
is generated by performing projective measurements on all the degrees of freedom in $B$. In Eq.~\ref{eq:PE-def}, $\varrho_A(z_B)$ is the state of $A$ given the measurement outcome $z_B$,
\eq{
\varrho_A(z_B) = \frac{1}{p(z_B)}{\rm Tr}_B[\Pi_{z_B}\varrho]\,;~~p(z_B) = {\rm Tr}[\Pi_{z_B}\varrho] \,,
\label{eq:PE-details}
}
where $\Pi_{z_B} = \ket{z_B}\bra{z_B}$ is a rank-1 projector and $p(z_B)$ is the probability of obtaining $z_B$ as the outcome. 
As an example, for a system of qubits, $z_B$ can be a bitstring $\{0,1\}^{\otimes |B|}$ corresponding to the measurement of the Pauli-$Z$ operator on each qubit of $B$.
The ensemble in Eq.~\ref{eq:PE-def} can also be fully characterised by all its moments with the $k^{\rm th}$ moment given by
\eq{
\rho_{{\cal E}_{\rm PE}}^{(k)} = \sum_{z_B}p(z_B) [\varrho_A(z_B)]^{\otimes k}\,.
\label{eq:moments-def}
}
Since the PE is induced on $A$ it is natural to consider it of finite size, ${|A|\sim O(1)}$, whereas the `measured bath', $B$ will be taken to be thermodynamically large, $|B|\gg 1$.

\begin{figure}[!t]
\begin{tikzpicture}[thick,scale=0.8, transform shape]
\draw (2.0,0) -- (2.0,1.8);
\draw (2.05,0) -- (2.05,1.8);
\draw(0.5,0) -- (0.5,2.2);
\draw(0.45,0) -- (0.45,2.2);
\node at (0., 3.2) {(a)};
\node at (3., 3.2) {(b)};
\node at (1.5, -0.9) {{\textcolor{white}{.}}};
\node at (0.5, 2.4) {$\rho_A(z_B)$};
\node at (2.0, 2.4) {$\Pi_{z_B}$};
\node at (0.65, 0.6) {$A$};
\node at (2.2, 0.6) {$B$}; 
\draw[fill=gray!20, rounded corners=3pt] (0,0) rectangle (2.5,0.4);
\node at (1.25, 0.2) {$\rho_0$};
\draw[fill=white, rounded corners=3pt] (0,0.8) rectangle (2.5,1.4);
\node at (1.25, 1.1) {$U_{\text{Haar}} \otimes U_{\text{Haar}}^*$};
\draw[fill=white] (1.7,1.8) rectangle (2.3,2.2);
\draw (1.83,1.95) arc (150:30:0.2);
\draw(2.0,1.85) -- (2.1,2.1);
\end{tikzpicture}
\hspace{-0.25cm}
\includegraphics[width=0.6\linewidth]{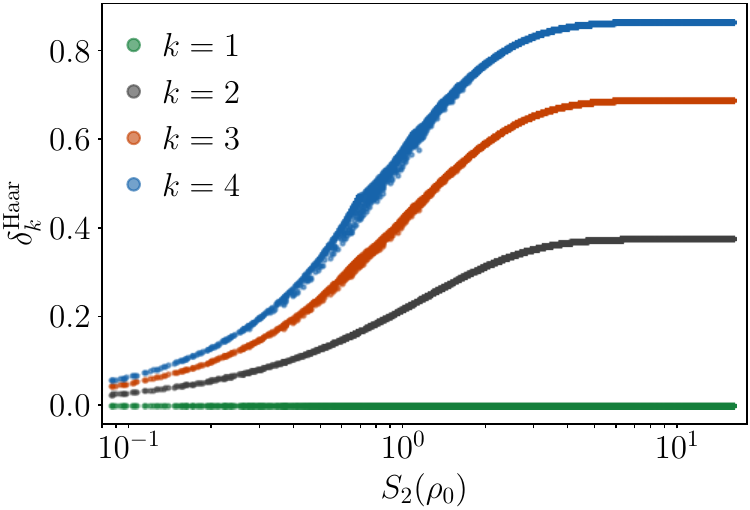}
    \caption{(a) Schematic showing the construction of $\eref[\rho_0]$, defined via Eq.~\ref{eq:Eref-def}. Note that $\rho_0$ itself is an operator and each black line carries two sets of indices, one from the kets and and other from the bras of $\rho_0$. The operators $U_{\rm Haar}$ and $U_{\rm Haar}^\dagger$ acting on them, respectively, have been folded into the one white box. (b) Trace distance between the $k^{th}$ moments of the $\mathcal{E}_\text{mix}$ and the Haar ensemble plotted against the second R{\'e}nyi entropy of the initial state for $|A|=2$.}
    \label{fig:eref}
\end{figure}

The central question of interest in this work is the fate of the PE in Eq.~\ref{eq:PE-def} when the corresponding $\varrho$ is obtained from evolving an initially mixed state with a chaotic unitary circuit.
We therefore need to define a reference ensemble which emerges for the PE following maximal scrambling or equivalently, by the action of an infinite-depth unitary circuit. 
Specifically, given an initial state $\rho_0$, the reference ensemble is defined as 
\eq{
\eref[\rho_0] := {\cal E}_{\rm PE}[U_{\rm Haar}\rho_0 U^\dagger_{\rm Haar}]\,,
\label{eq:Eref-def}
}
where $U_{\rm Haar}$ is a Haar-random unitary operator acting on the entire system $A\cup B$; the construction of ${\eref}$ is depicted graphically in Fig.~\ref{fig:eref}(a).
The definition, \ref{eq:Eref-def}, is guided by two rationales.
First, in the limit of $\rho_0$ pure, $\eref[\rho_0]$ would necessarily deep thermalise to the Haar ensemble on $A$.
Hence any deviation from the Haar ensemble or deep thermalisation altogether for a generic mixed $\rho_0$ can be attributed solely to its mixedness. 
Secondly, as discussed above, $U_{\rm Haar}\rho_0 U^\dagger_{\rm Haar}$ represents statistically a maximally scrambled state attainable from $\rho_0$.
Consequently, the distance of $\eref[\rho_0]$ from the Haar ensemble is expected to provide a bound on the distance from the Haar ensemble of the PE induced on a state resulting from evolving $\rho_0$ via any finite-depth unitary circuit. 

With the definition of $\eref[\rho_0]$ firmly in place, we address the question that in what sense is it (or not) a deep thermal ensemble by computing analytically its moments defined in Eq.~\ref{eq:moments-def} in the limit of $|B|\to\infty$.
{\news{Importantly, any moment depends solely on the eigenvalues of $\rho_0$. While the $k^{\rm th}$ moment is given explicitly in the Supplementary Material~\cite{supp} it will be sufficient to discuss $k=1,2$ here.}}

The first moment, $k=1$, is trivially the maximally mixed state, $\rho^{(k=1)}_{\eref[\rho_0]}=D_A^{-1}\mathbb{I}_{D_A}$.
For $k=2$, the first non-trivial moment, it reduces to
\eq{
\rho^{(k=2)}_{\eref[\rho_0]}=\frac{\mathbb{I}_{D_A^2}+({\rm Tr}\rho_0^2)~\mathbb{S}_{D_A^2}}{D_A^2+({\rm Tr}\rho_0^2)~D_A}\,,
\label{eq:eref-second-moment}
}
where ${\mathbb{S}}_{D_A^2}$ is the $\mathsf{SWAP}$ operator acting on the doubled Hilbert space ${\cal H}_A^{\otimes 2}$.
An important corollary of the result is that the average purity of the ensemble is 
\eq{
{\cal P}_{\eref[\rho_0]} = {\rm Tr}\left[\rho^{(2)}_{\eref[\rho_0]}{\mathbb{S}}_{D_A^2}\right]=\frac{1+D_A{\rm Tr}\rho_0^2}{D_A+{\rm Tr}\rho_0^2}~\leq 1\,,
\label{eq:purity-eref}
}
with the equality appearing only for $\rho_0$ pure.
The above inequality implies that for any finite mixedness in $\rho_0$, irrespective of however small, the average purity ${\cal P}_{\eref[\rho_0]}<1$ necessarily. 
Therefore, $\eref[\rho_0]$ cannot be described by any ensemble of pure states, which automatically rules out the Haar or a Scrooge ensemble. 
This leads to us to the first key result of this work -- {\it the emergence of exact quantum designs via PEs generated from scrambling dynamics is fine-tuned to pure initial states; any finite entropy in the initial state leads to a finite departure of the PE from a design}~\footnote{It can be shown that the trace distance between the $k^{\rm th}$ moments $\eref[\rho_0]$ and the Haar ensemble is bounded from below by a function $f[S_2(\rho_0)]$ which grows linearly with $S_2(\rho_0)$ for $S_2(\rho_0)\ll 1$.}.
To put this on a quantitative footing, we compute the distance of the $k^{\rm th}$-moment of $\eref[\rho_0]$ and that of the Haar ensemble,
$\delta_k^{\rm Haar} = \frac{1}{2}\big|\big|\rho_{\eref[\rho_0]}^{(k)}-\rho_{\rm Haar}^{(k)}\big|\big|_1$.
In Fig.~\ref{fig:eref}(b), we show the results for $\delta_k^{\rm Haar}$ as a function of the second R\'enyi entropy of the initial state, $S_2(\rho_0) = -\ln{\rm Tr}[\rho_0^2]$ which clearly shows that the distance between $\eref[\rho_0]$ and the Haar ensemble grows monotonically with $S_2(\rho_0)$ for $k\ge 2$.

Having established that $\eref[\rho_0]$ does not correspond to any ensemble of pure states, a natural question arises that can it be described by a `universal' ensemble of mixed states.
In particular, one can ask does there exist a maximum entropy ensemble of pure states on $A$ augmented by auxiliary degrees of freedom, $X$, such tracing out $X$ in the states in the ensemble yields $\eref[\rho_0]$.
To address this question, consider a purification of $\rho_0$ via state $\ket{\phi_0}$ on an extended system $X\cup A\cup B$, where the dimension of $X$ is the same as that of $A\cup B$.
The state is most conveniently represented diagrammatically as,
\eq{
\ket{\phi_0} = 
\cbox{
\begin{tikzpicture}[thick]
\draw (0.25,-0.15) -- (0.25,0.75);
\draw (0.75,-0.25) -- (0.75,0.75);
\draw (0.25,-0.15) arc (0:-90:0.1);
\draw (0.75,-0.25) arc (0:-90:0.1);
\draw (0.15,-0.25) -- (-0.15,-0.25);
\draw (0.65,-0.35) -- (-0.25,-0.35);
\draw (-0.15,-0.25) arc (270:180:0.1);
\draw (-0.25,-0.35) arc (270:180:0.1);
\draw (-0.25,-0.15) -- (-0.25,0.75);
\draw (-0.35,-0.25) -- (-0.35,0.75);
\draw[fill=gray!20, rounded corners=3pt] (0,0) rectangle (1,0.5);
\node at (0.5, 0.25) {$\sqrt{\rho_0}$};
\node at (-0.125, 0.875) {$X$};
\node at (0.375, 0.875) {$A$};
\node at (0.875, 0.875) {$B$};
\end{tikzpicture}}=
\cbox{
\begin{tikzpicture}[thick]
\draw (0.25,0) -- (0.25,0.25);
\draw (0.75,0) -- (0.75,0.25);
\draw (-0.25,0) -- (-0.25,0.25);
\draw (-0.35,0) -- (-0.35,0.25);
\node at (-0.125, 0.375) {$X$};
\node at (0.375, 0.375) {$A$};
\node at (0.875, 0.375) {$B$};
\filldraw[fill=gray!30] (-0.45,0) -- (0.95,0) -- (0.25,-0.25) -- cycle;
\end{tikzpicture}}\,,
}
which makes it trivial to see that $\rho_0 = {\rm Tr}_X[\ket{\phi_0}\bra{\phi_0}]$.
Applying the Haar-random unitary on $A\cup B$ to the state $\ket{\phi_0}$ and performing measurements on $B$ leads to a PE of pure states on $X\cup A$, ${\cal E}_{\rm PE}^{XA}=\{p(z_B),\ket{\phi_{XA}(z_B)}\}_{z_B}$, where $\ket{\phi_{XA}(z_B)}$ is given by
\eq{
\ket{\phi_{XA}(z_B)} = 
\cbox{
\begin{tikzpicture}[thick]
\draw (0.25,0) -- (0.25,1.25);
\draw (0.75,0) -- (0.75,1.25);
\draw (-0.25,0) -- (-0.25,1.25);
\draw (-0.35,0) -- (-0.35,1.25);
\filldraw[fill=gray!30] (-0.45,0) -- (0.95,0) -- (0.25,-0.25) -- cycle;
\draw[fill=white, rounded corners=3pt] (0,0.25) rectangle (1,1.0);
\node at (0.5, 0.625) {$U_{\text{Haar}}$};
\draw[fill=white] (0.55,1.25) rectangle (0.95,1.5);
\draw (0.62,1.35) arc (150:30:0.15);
\draw (0.75,1.3) -- (0.8,1.47);
\node at (1.2, 1.375) {$z_B$};
\end{tikzpicture}}\,.
}
$\eref[\rho_0]$ can be obtained from ${\cal E}_{\rm PE}^{XA}$ as
\eq{
\eref[\rho_0] = \{p(z_B), {\rm Tr}_X\ket{\phi_{XA}(z_B)}\bra{\phi_{XA}(z_B)}\}_{z_B}\,.
}
The $k^{\rm th}$-moment is then given by~\cite{supp} 
\eq{
\rho_{\eref[\rho_0]}^{(k)} = &\sum_{z_B}p(z_B)\left[{\rm Tr}_X\ket{\phi_{XA}(z_B)}\bra{\phi_{XA}(z_B)}\right]^{\otimes k}\nonumber\\
=&\!\!\int \!\!d\psi_{\rm Haar}  D_{XA}  \frac{\left({\rm Tr}_X [\sqrt{\rho_{XA}}\ket{\psi}\!\!\bra{\psi}\sqrt{\rho_{XA}}]\right)^{\otimes k}}{\braket{\psi|\rho_{XA}|\psi}^{k-1}}\,,
\label{eq:eref-scrooge-res}
}
where $d\psi_{\rm Haar}$ is the Haar-measure over the Hilbert space of $X\cup A$ (of dimension $D_{XA}$) and  $\rho_{XA} = (\rho_0^T \otimes \mathbb{I}_{D_A})/D_A$.

The most important fallout of the result in Eq.~\ref{eq:eref-scrooge-res} is that it answers the question raised at the beginning of this discussion in the affirmative -- 
{\it indeed $\eref[\rho_0]$ can be interpreted as an ensemble obtained by tracing out auxiliary degrees of freedom, $X$, from a maximum entropy ensemble, in this case the Scrooge ensemble, defined by the density matrix $\rho_{XA}$ on $X\cup A$; in this sense $\eref[\rho_0]$ can be dubbed as a mixed state deep thermal ensemble.}
A key distinction from deep thermal ensembles of pure states is that constructing $\eref[\rho_0]$ as a partial trace from a maximum-entropy ensemble requires the full spectrum of $\rho_0$, a mixed state on a thermodynamically large system even though $\eref[\rho_0]$ itself is defined on a finite subsystem. In contrast, for pure states, the (finite-dimensional) density matrix of the subsystem often suffices to define the induced deep-thermal ensemble.

An interesting case of the above constructions is when all the non-zero eigenvalues of $\rho_0$ are the same and equal to $1/{\rm rank}(\rho_0)$.
In this case, $\eref[\rho_0]$ is  given by tracing out $R$ from a Haar ensemble defined on $R\cup A$~\cite{supp},
\eq{
\eref[\rho_0]=\{{\rm Tr}_R[\ket{\psi}\bra{\psi}]:\ket{\psi}\sim {\rm Haar}(D_{RA})\}\,,
\label{eq:ghse}
}
where the dimension of $R$ is set by ${D_R = {\rm rank}(\rho_0)}$.
The ensemble in Eq.~\ref{eq:ghse} is referred to as the generalised Hilbert-Schmidt ensemble (gHSe)~\cite{hall1998randomquantumcorrelations,karol2001inducedmeasures,bansal2025pseudordm}.
The gHSe was shown~\cite{yu2025mixedstatedeepthermalization} to emerge as the PE for lossy measurements in a fine-tuned situation with a specific choices of unitary, measurement, and initial state.
Our result lays down the general conditions in which the PE is described by an appropriate gHSe.

\begin{figure*}
\adjustbox{valign=c}{
\begin{tikzpicture}[thick, scale=0.62, every node/.style={font=\footnotesize}]
\draw[rounded corners=3pt,fill=Purple!10,draw=Purple,dashed] (-0.3,-0.25) rectangle (8.2,0.75);
\node at (-0.5, 5) {(a)};
\foreach \x in {0,...,8}{
\draw[color=gray](\x,-0.45) -- (\x,4);
}
\foreach \x in {0,...,8}{
\foreach \y in {0,...,3}{
\draw[fill=gray,draw=gray] (1*\x,1*\y) circle (0.08);
\draw[color=gray](0,\y) -- (8,\y);
\node[draw, fill=RoyalBlue!20,draw=gray, diamond, minimum size=6,inner sep=0pt] at (\x,\y+0.5) {};
}}
\foreach \x in {0,...,7}{
\foreach \y in {0,...,3}{
\draw[fill=Orange!20,draw=gray] (\x+0.5-0.1,\y-0.1) rectangle (\x+0.5+0.1,\y+0.1);
}}
\draw[rounded corners=3pt, fill=gray!20,draw=gray] (-0.2,-1.05) rectangle (3.2,-0.45);
\foreach \x in {4,...,8}{
\filldraw[fill=blue!20] (\x-0.2,-0.45) -- (\x+0.2,-0.45) -- (\x,-0.65) -- cycle;
\filldraw[fill=red!20] (\x-.2,4) -- (\x+0.2,4) -- (\x,4.2) -- cycle;
}
\foreach \x in {2,3}{
\filldraw[fill=red!20] (\x-.2,4) -- (\x+0.2,4) -- (\x,4.2) -- cycle;
}
\def\eps{0.1}
\foreach \x in {0,...,3}{
\draw[color=black](\x,-1.05) -- (\x,-1.35);
\draw[color=black](\x-\eps,-1.05-\eps) -- (\x-\eps,-1.35-\eps);
\draw(\x,-1.35) to[out=-90,in=-90] (\x-\eps,-1.35-\eps);
}
\foreach \x in {0,...,8}{
\draw(\x-\eps,-0.45-\eps) -- (\x-\eps,4-\eps);
}
\foreach \x in {0,...,8}{
\foreach \y in {0,...,3}{
\draw[fill=black] (1*\x-\eps,1*\y-\eps) circle (0.1);
\draw(0-\eps,\y-\eps) -- (8-\eps,\y-\eps);
\node[draw, fill=RoyalBlue,diamond, minimum size=6,inner sep=0pt] at (\x-\eps,\y+0.5-\eps) {};
}}
\foreach \x in {0,...,7}{
\foreach \y in {0,...,3}{
\draw[fill=Orange] (\x+0.5-0.1-\eps,\y-0.1-\eps) rectangle (\x+0.5+0.1-\eps,\y+0.1-\eps);
}}
\draw[rounded corners=3pt, fill=black!20] (-0.2-\eps,-1.05-\eps) rectangle (3.2-\eps,-0.45-\eps);
\node at (1.5-\eps, -0.75-\eps) {$\sqrt{\rho_S}$};
\foreach \x in {4,...,8}{
\filldraw[fill=blue!20] (\x-0.2-\eps,-0.45-\eps) -- (\x+0.2-\eps,-0.45-\eps) -- (\x-\eps,-0.65-\eps) -- cycle;
\filldraw[fill=red!20] (\x-.2-\eps,4-\eps) -- (\x+0.2-\eps,4-\eps) -- (\x-\eps,4.2-\eps) -- cycle;
}
\foreach \x in {2,3}{
\filldraw[fill=red!20] (\x-.2-\eps,4-\eps) -- (\x+0.2-\eps,4-\eps) -- (\x-\eps,4.2-\eps) -- cycle;
}
\draw[dashed,Bittersweet](1.3,-0.45)--(1.2,4.3);
\draw [thick, decorate, decoration={brace,amplitude=5pt,mirror,raise=4pt},yshift=0pt]
(8.1,0) -- (8.1,3.5) node [black,midway,xshift=15pt] {$t$};
\draw [thick, decorate, decoration={brace,amplitude=5pt, raise=3pt},yshift=0pt]
(-0.1,4.1) -- (1.1,4.1) node [black,midway,yshift=15pt] {$A$};
\draw [thick, decorate, decoration={brace,amplitude=5pt, raise=3pt},yshift=0pt](2,4.1) -- (8,4.1) node [black,midway,yshift=15pt] {$B$};
\draw [thick, decorate, decoration={brace,mirror,amplitude=5pt, raise=3pt},yshift=0pt](3.7,-0.7) -- (8.1,-0.7) node [black,midway,yshift=-15pt] {$E$};
\draw [thick, decorate, decoration={brace,mirror,amplitude=5pt, raise=3pt},yshift=0pt](-0.1,-1.35) -- (3.1,-1.35) node [black,midway,yshift=-15pt] {$S$};
\end{tikzpicture}
}
\adjustbox{valign=c}{
\includegraphics[width=0.35\linewidth]{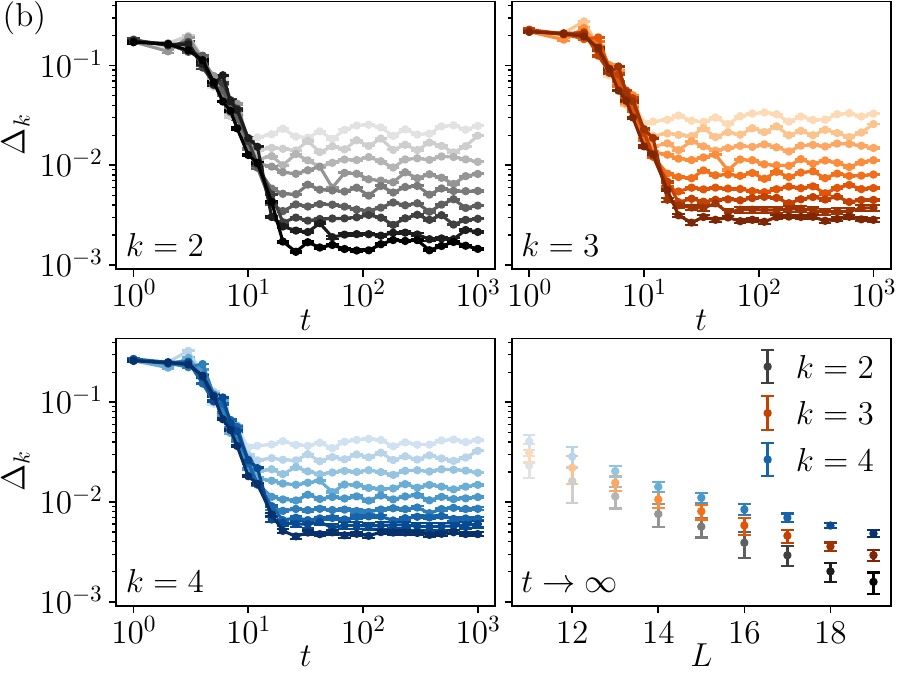}
}
\adjustbox{valign=c}{
\includegraphics[width=0.25\linewidth]{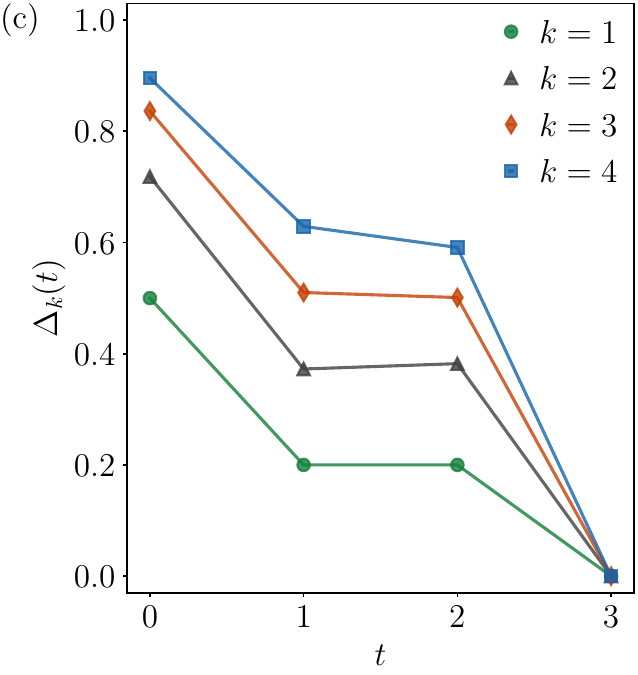}
}
\caption{(a) The kicked Ising chain in Eq.~\ref{eq:UF-kim} as a circuit with the shaded box denoting $U_F$. The two copies represent $U_t$ and $U_t^\dagger$. The red triangles on subsystem $B$ denote projective measurements in the computational basis whereas the blue triangles on $E$ along with the mixed state on $S$ denotes the initial state in Eq.~\ref{eq:rho0-kim}. (b) Results for $\Delta_k(t)$ (defined in Eq.~\ref{eq:Delta}) for $(J,g,h)=(0.8,0.6472,0.7236)$. Different panels correspond to different values of $k$. The initial state is of the form in Eq.~\ref{eq:rho0-kim} with $|S|=3$. $|A|=2$ is held fixed and different intensities correspond to different $|B|=9,10...17$ with darker colours denoting higher values. The data is averaged over $100$ initial states. (c) Results for $\Delta_k(t)$ at a self-dual point $(J,g,h)=(\pi/4,\pi/9,\pi/4)$, with $|S|=1$ and $\ket{v_j}=\ket{+}$ (see Eq.~\ref{eq:rho0-kim}) in which case the moments can be computed analytically in the limit of $|B\cap E|\to\infty$.}
\label{fig:kim}
\end{figure*}

This concludes our characterisation of $\eref[\rho_0]$ for $\rho_0$ mixed and discussion of how can it be interpreted as a deep thermal ensemble. 
We next discuss how $\eref[\rho_0]$ emerges naturally in locally interacting chaotic systems. 
In particular, we consider a non-integrable periodically kicked (Floquet) Ising chain for which we show numerical results and its self-dual point, we analytically prove the emergence of $\eref[\rho_0]$ at finite times.

For a Floquet system, the time-evolution operator for time $t \in {\mathbb{Z}}$ is given by $U_t = U_F^t$ with $U_F$ the time-evolution operator over one period, often referred to as the Floquet unitary.
For the kicked Ising chain, 
\eq{
U_F = U_Y U_Z = e^{-ih\sum_j Y_j}e^{-i\sum_j [JZ_jZ_{j+1} + gZ_j]}\,,
\label{eq:UF-kim}
}
with $Y_j (Z_j)$ the Pauli-$Y (Z)$ operator at site $j$, and we consider open boundary conditions.
The model can be represented as a circuit, as in Fig.~\ref{fig:kim}(a) where 
$\cbox{\begin{tikzpicture}[thick,scale=0.65]
    \draw[fill=black] (-0.5,0) circle (0.1);
    \draw[fill=black] (0.5,0) circle (0.1);
    \draw(-0.5,0) -- (0.5,0);
    \draw(-0.5,-0.2) -- (-0.5,0.2);
    \draw(0.5,-0.2) -- (0.5,0.2);
    \draw[fill=Orange] (-0.1,-0.1) rectangle (0.1,0.1);
\end{tikzpicture}}$
denotes the Ising interaction and the longitudinal fields in $U_Z$ and $\cbox{\begin{tikzpicture}[thick,scale=0.65]
    \draw(-0.5,0) -- (0.5,0);
    \node[draw, fill=RoyalBlue,diamond, minimum size=5,inner sep=0pt] at (0,0) {};
\end{tikzpicture}}$,
the transverse kicks in $U_Y$.
We consider initial states of the form (see Fig.~\ref{fig:kim}(a))
\eq{
\rho_0 = \rho_S\otimes \prod_{j \in E}\ket{v_j}\bra{v_j}\,,
\label{eq:rho0-kim}
}
where $\rho_S$ is an arbitrary mixed state on the $S$ leftmost qubits and $\ket{v_j}$ is a random pure state of the qubit at $j\in E$ with $E=\overline{S}$. 
Evolution of $\rho_0$ with $U_F$ for time $t$ yields $\rho_t=U_F^t\rho_0 U_F^{\dagger t}$ from which we generate the PE, ${\cal E}_{\rm PE}[\rho_t]$ on the leftmost $|A|$ qubits, by performing projective measurements on $B$ in the computational basis.
To quantify the distance between ${\cal E}_{\rm PE}[\rho_t]$ and $\eref[\rho_0]$ with $t$ we employ the trace distance between the $k^{\rm th}$ moments 
\eq{
\Delta_k(t) = \frac{1}{2}\big|\big|\rho_{\eref[\rho_0]}^{(k)} - \rho_{{\cal E}_{\rm PE}[\rho_t]}^{(k)} \big|{\big|}_1\,.
\label{eq:Delta}
}
In Fig.~\ref{fig:kim}(b), we show numerical results for $\Delta_k(t)$ for generic values of $(J,g,h)$ for a few values of $k$; the results show concomitantly that in the limit of $|B|\to\infty$ and $|A|\sim O(1)$, for all the considered moments, $\Delta_k(t\to\infty)\to 0$.
This provides strong evidence that the moments of ${{\cal E}_{\rm PE}[\rho_t]}$ approach those of $\eref[\rho_0]$ at late times.
This signals the emergence of a deep thermal ensemble from a mixed initial state evolving under the kicked Ising chain, but in the limit of $t\to\infty$.
\news{The results for the timescales and finite-size scaling are relegated to Ref.~\cite{supp}.}

We next turn to the self-dual point of the model, $J=\pi/4=h$, where 
$\cbox{\begin{tikzpicture}[thick,scale=0.65]
    \draw(-0.5,0) -- (0.5,0);
    \node[draw, fill=RoyalBlue,diamond, minimum size=5,inner sep=0pt] at (0,0) {};
\end{tikzpicture}} = \cbox{\begin{tikzpicture}[thick,scale=0.65]
    \draw(-0.5,0) -- (0.5,0);
    \node[draw, fill=white,diamond, minimum size=5,inner sep=0pt] at (0,0) {};
\end{tikzpicture}}=\cbox{\begin{tikzpicture}[thick,scale=0.65]
    \draw(-0.5,0) -- (0.5,0);
    \draw[fill=Orange] (-0.1,-0.1) rectangle (0.1,0.1);
\end{tikzpicture}}/\sqrt{2}$ upto an irrelevant global phase with 
$\cbox{\begin{tikzpicture}[thick,scale=0.65]
    \draw(-0.5,0) -- (0.5,0);
    \node[draw, fill=white,diamond, minimum size=5,inner sep=0pt] at (0,0) {};
\end{tikzpicture}}$ denoting the 
Hadamard gate.
This implies that the self-dual kicked Ising (SDKI) chain is a special instance of a dual-unitary circuit, which has enabled a range of exact results for the model (see Ref.~\cite{bertini2025exactlysolvablemanybodydynamics} for a review and further references therein).
Notably, for a pure initial state of the form $\otimes \prod_j(\ket{+}_j\bra{+})$, where $\ket{+}$ denotes an $X$-polarised state, evolution under the SDKI leads to the exact emergence of unitary designs: the PE coincides with the Haar ensemble exactly for $t \geq |A|$~\cite{ho2022exact}.
As we discuss below, the exact solvability continues to hold for mixed initial states of the form in Eq.~\ref{eq:rho0-kim} provided $\ket{v_j}=\ket{+}~\forall j \in E$.

It is useful to consider a purification of $\rho_0$ via a state $\ket{\phi_0}$ on an extended system $X\cup S\cup E$ where $|X| = |S|$.
The SDKI unitary acting on $S\cup E$, followed by measurements on $B$ induce a PE of pure states on $X\cup A$, ${\cal E}_{\rm PE}^{XA}=\{p(z_B),\phi_{XA}(z_B)\}$,
where $\ket{\phi_{XA}(z_B)}$ is given by
\eq{
\ket{\phi_{XA}(z_B)} =\!\!\! 
\cbox{
\begin{tikzpicture}[thick]
\draw (0.25,-0.6) -- (0.25,1.25);
\draw (1.0,0) -- (1.0,1.25);
\draw (-0.25,-.6) -- (-0.25,1.25);
\draw (-0.25,-.6) arc (180:270:0.1);
\draw (0.25,-.6) arc (0:-90:0.1);
\draw (-0.15,-.7) -- (0.15,-.7);
\filldraw[fill=Blue!30] (0.85,0) -- (1.15,0) -- (1.0,-0.15) -- cycle;
\draw[fill=white, rounded corners=3pt] (0,0.25) rectangle (1.25,1.0);
\draw[fill=gray!20, rounded corners=3pt] (-0.1,-0.5) rectangle (0.6,0);
\node at (0.625, 0.625) {$U^t_{\rm SDKI}$};
\node at (0.25, -0.25) {$\sqrt{\rho_S}$};
\draw[fill=white] (0.55+0.25,1.25) rectangle (0.95+0.25,1.5);
\draw (0.62+0.25,1.35) arc (150:30:0.15);
\draw (0.75+0.25,1.3) -- (0.8+0.25,1.47);
\node at (1.2+0.25, 1.375) {$z_B$};
\node at (0.25, 1.4) {$A$};
\node at (-0.25, 1.4) {$X$};
\node at (1, -0.4) {$E$};
\end{tikzpicture}}
\!\!\!\!\!\!\overset{|B\cap E|\to\infty}{\asymp}\!\!\!\!\!\!
\cbox{
\begin{tikzpicture}[thick]
\draw (0.25,-0.6) -- (0.25,1.25);
\draw (1.0,0.5) -- (1.0,1.25);
\draw (-0.25,-.6) -- (-0.25,1.25);
\draw (-0.25,-.6) arc (180:270:0.1);
\draw (0.25,-.6) arc (0:-90:0.1);
\draw (-0.15,-.7) -- (0.15,-.7);
\draw[fill=white,rounded corners=3pt] (0,0.25) rectangle (1.25,1.0);
\draw(1.25,0.625)--(1.45,0.625);
\filldraw[fill=Green!10] (1.45,0.3) -- (1.45,0.9) -- (1.75,0.6) -- cycle;
\node at (1.625, 0) {$\psi_{\rm Haar}^{(2^t)}$};
\draw[fill=gray!20, rounded corners=3pt] (-0.1,-0.5) rectangle (0.6,0);
\node at (0.625, 0.625) {$U^t_{\rm SDKI}$};
\node at (0.25, -0.25) {$\sqrt{\rho_S}$};
\draw[fill=white] (0.55+0.25,1.25) rectangle (0.95+0.25,1.5);
\draw (0.62+0.25,1.35) arc (150:30:0.15);
\draw (0.75+0.25,1.3) -- (0.8+0.25,1.47);
\node at (1.2+0.45, 1.375) {$z_{B\cap S}$};
\node at (0.25, 1.4) {$A$};
\node at (-0.25, 1.4) {$X$};
\end{tikzpicture}}.
\label{eq:SDKI-haar}
}
The statistical equivalence in the above equation implies that, in the limit of $|B\cap E|\to\infty$, for the SDKI chain with the solvable initial states, the states in ${\cal E}_{\rm PE}^{XZ}$ samples the ensemble of states obtained replacing the SDKI circuit acting on $B\cap E$ by a Haar random state (of dimension $2^t$) in the temporal direction~\cite{ho2022exact}.
This crucial simplification, which is a feature of the SDKI chain, is key towards the its solvability for the class of initial mixed states discussed above. 
Using the standard toolbox of Haar averaging~\cite{weingarten1978asymptotic} and the dual-unitarity of the SDKI chain, the moments of the ${\cal E}_{\rm PE}^{X\cup A}$, comprised of states in Eq.~\ref{eq:SDKI-haar}, can be computed explicitly~\cite{supp}.
By tracing out the (copies of) $X$ from the said moments, the moments of ${\cal E}_{PE}^{A}[\rho_0]$ can be, in turn, computed explicitly as well. For $t\geq |A|+|S|$, we find
\eq{
 \rho^{(k)}_{{\cal E}_{\rm PE}^A[\rho_t]}=&\!\!\int \!\!d\psi_{\rm Haar}  D_{XA}  \frac{\left({\rm Tr}_X [\sqrt{\rho_{XA}}\ket{\psi}\!\!\bra{\psi}\sqrt{\rho_{XA}}]\right)^{\otimes k}}{\braket{\psi|\rho_{XA}|\psi}^{k-1}}\,,
 \label{eq:SDKI-rho-k}
}
where $\rho_{XA}=(\rho^T_S\otimes \mathbb{I}_{D_A})/D_A$. This is exactly analogous to Eq.~\ref{eq:eref-scrooge-res}, which demonstrates that for $t\geq |A|+|S|$, the moments of ${\cal E}_{\rm PE}^A[\rho_t]$ are exactly equal to those of $\eref[\rho_S]$. 
In other words, the moments of ${\cal E}_{\rm PE}^A[\rho_t]$ can be written as partial traces of the moments of a Scrooge ensemble defined on an extended system, and this emerges at a finite time $t_{\rm ref} = |A|+|S|$.
It therefore implies, that for a class of mixed initial states, dynamics with the SDKI chain does lead to the emergence of a mixed-state deep thermal ensemble on subsystem $A$ at a finite time.
While we can analytically show that Eq.~\ref{eq:SDKI-rho-k} hold for $t\geq |A|+|S|$, our numerical results suggest that the bound is optimal and indeed 
$
\big|\big| \rho_{\eref[\rho_S]}^{(k)} - \rho_{{\cal E}_{\rm PE}[\rho_{t=|A|+|S|}]}^{(k)}\big|\big|_{1}\to 0\,.
$
An exemplary result is shown in Fig.~\ref{fig:kim}(c).
This concludes our discussion of the kicked Ising chain.

\news{Numerical results for Floquet mixed-field Ising chain~\cite{supp} also shows identical phenomenology, establishing generality.}

We now close with a brief summary and some future outlook.
In this work we investigated the fate of deep thermalisation under unitary dynamics of an initially mixed state, $\rho_0$.
We analytically compute the moments of the PE, $\eref[\rho_0]$, on a state obtained by evolving $\rho_0$ under a generic Haar-random unitary.
The results show that the notions of deep thermalisation as defined for PEs of pure states, such as emergence of quantum designs for dynamics without any conservation laws, or more generically, the emergence of maximum entropy ensembles subject to constraints placed by conservations laws are fragile to any amount of mixedness in the initial state.
This implies that the phenomenology of deep thermalisation of mixed states under unitary dynamics is fundamentally different from that of pure states -- this calls for a reformulation of the maximum entropy principle for mixed states.
To address this, we show that $\eref[\rho_0]$ can be understood as a partially traced out version of a maximum entropy ensemble induced on a purification of the state on a system augmented with auxiliary degrees of freedom. 
The maximum entropy ensemble on the augmented system is described by a Scrooge ensemble defined by $\rho_0$ -- this concretely defines a new paradigm for deep thermalisation of mixed states and constitutes the first main result of this work.
An important conceptual point of departure from deep thermalisation for pure states is that, for mixed states the maximum entropy ensemble is defined with respect to a density matrix on a thermodynamically large system whereas for pure states the reduced density matrix of the local subsystem suffices.

While $\eref[\rho_0]$ is defined with reference to a Haar-random unitary, we show that it emerges robustly for generic locally interacting, chaotic dynamics.
We demonstrate this numerically using a non-integrable kicked Ising chain and exploit the exact solvability (due to dual-unitarity) at its self-dual point to analytically prove the emergence of $\eref[\rho_0]$ at finite times. This constitutes the second main result of this work.

The general idea of deep thermalisation of mixed states is quite nascent.
It is therefore natural that there several open questions of immediate interest, some of them raised by this work as well.
An obvious extension of this work will be consider dynamics with conservation laws and its interplay with the extent to which $\rho_0$ breaks the conservation laws and to which, the measurements reveal   the conserved charge. 
{\new {Similarly, it will be interesting to understand fate of our results in ergodicity-broken systems. In fact, our preliminary results~\cite{supp} suggest that mixed-state deep thermalisation measures can serve as a sensitive probe for weak ergodicity breaking.
}}

Within the realms of the exactly solvable SDKI chain, in our case the mixedness in the initial state was confined to a contiguous subsystem. It will be interesting to study the fate of deep thermalisation and associated timescales when the mixedness is sprinkled throughout the system.
In addition, it will be interesting to develop solvable models which show distinct timescales for different moments of the PE exhibiting deep thermalisation, along the lines developed for pure states~\cite{ippoliti2022solvablemodelofdeep}.

{\new{

\news{Pertinently, both decoherence and unitary scrambling can produce featureless reduced states (${\rho_A\sim\mathbb{I}_{D_A}}$) which cannot be distinguished by conventional thermalisation. However, deep-thermalisation measures, by probing higher moments can do so, and therein lies the relevance of it, especially for mixed states}
Crucially, these signatures are experimentally accessible. Observables based on probabilities of probabilities (PoPs)~\cite{shaw2024universal} and fidelity estimators~\cite{choi2023preparing} provide natural probes. Our preliminary results~\cite{supp} demonstrate that PoPs over $\eref$ follow universal Erlang distributions -- sharply distinct from the Porter–Thomas or $\delta$-function statistics of pure-state designs or featureless ensembles, respectively -- offering a direct operational handle on mixed-state deep thermalisation.

}}

\stoptoc

\begin{acknowledgments}
We thank P. W. Claeys, S. Mandal, S. Manna and G. J. Sreejith for useful discussions and collaborations on related work. This work was supported by the Department of Atomic Energy, Government of India under Project No. RTI4001, by SERB-DST, Government of India under Grant No. SRG/2023/000858 and by a Max Planck Partner Group grant between ICTS-TIFR, Bengaluru and MPIPKS, Dresden.
\end{acknowledgments}


\bibliography{refs}

\resumetoc

\clearpage

\setcounter{equation}{0}
\setcounter{figure}{0}
\setcounter{page}{1}
\renewcommand{\theequation}{S\arabic{equation}}
\renewcommand{\thefigure}{S\arabic{figure}}
\renewcommand{\thesection}{S\arabic{section}}
\renewcommand{\thepage}{S\arabic{page}}
\newcommand\imat{\begin{bmatrix}
    \mathbb{I} & i\mathbb{I}
\end{bmatrix}}
\newcommand\timat{\begin{bmatrix}
    \mathbb{I} \\ -i\mathbb{I}
\end{bmatrix}}

\onecolumngrid
\begin{center}
{\large{\bf Supplementary Material: {\mytitle}}}\\
\medskip

Alan Sherry and Sthitadhi Roy\\
{\small{\it International Centre for Theoretical Sciences, Tata Institute of Fundamental Research, Bengaluru 560089, India}}
\end{center}
\bigskip

\tableofcontents

\section{Computation of the moments of $\eref[\rho_0]$}
\label{sec:mom-comp}
In the main text, we mentioned that the $k^{\rm th}$ moment of $\eref[\rho_0]$ depends only on the eigenvalues of $\rho_0$. It is given explicitly as~\footnote{A technical point to note here is that in the limit of $|B|\to\infty$, the result in Eq.~\ref{eq:eref-moments} holds for a single Haar-random unitary~\cite{supp}},
\eq{
\rho^{(k)}_{\eref[\rho_0]}=\frac{\sum_{\sigma \in \mathcal{S}_k} h_\sigma(\rho_0) \ \text{Perm}_{\mathcal{H}_A^{\otimes k}}(\sigma)}{\sum_{\sigma \in \mathcal{S}_k} h_\sigma(\rho_0)D_A^{|\sigma|}}\,,
\label{eq:eref-moments}
}
where $\text{Perm}_{\mathcal{H}_A^{\otimes k}}(\sigma)$ permutes the $k$ copies of the $D_A$-dimensional Hilbert space of $A$, $\mathcal{H}_A$, according to an element $\sigma$ in
the permutation group of $k$ elements $\mathcal{S}_k$. The dependence of the moment on $\rho_0$ via its eigenvalues is in $h_\sigma(\rho_0) \in [0,1]$, given by 
\begin{equation}
h_\sigma(\rho_0)=\text{Tr}\rho_0^{n_1}...\text{Tr}\rho_0^{n_{|\sigma|}}=\sum_{l_1...l_{|\sigma|}}\lambda_{l_1}^{n_1}...\lambda_{l_{|\sigma|}}^{n_{|\sigma|}}\,,  \label{eq:hsigma}
\end{equation}
where $|\sigma|$ is the number of cycles and $n_m$ is the order of the $m^{th}$ cycle of $\sigma$ with $\sum_{m=1}^{|\tau|}n_m=k$. Here $\{\lambda_i\}$ denotes the set of eigenvalues of $\rho_0$.

Before we detail the derivations of Eq.~\ref{eq:eref-moments} and Eq.~\ref{eq:hsigma}, we discuss some pertinent cases. 
Two trivial limits are (i) $\rho_0$ pure where $h_\sigma(\rho_0)=1~\forall \sigma$ in which case $\eref[\rho_0]$ reduces to the Haar ensemble as expected~\cite{cotler2023emergent}, and (ii) $\rho_0\propto \mathbb{I}$,; in this case $h_\sigma(\rho_0) = \delta_{e\sigma}$ where $\sigma$ is the identify element of ${\cal S}_k$ where moments of $\eref[\rho_0]$ are trivially $\propto \mathbb{I}^{\otimes k}$. 

We now derive Eq.~\ref{eq:eref-moments} and Eq.~\ref{eq:hsigma} in detail. Given the initial state $\rho_0$ (of a $d$-dimensional Hilbert space), the unitarily evolved states are of the form $U\rho_0 U^\dagger$, where $U\sim\text{Haar}(d)$. The projected ensemble $\eref[\rho_0]:=\{p(z_B),\rho_A(z_B)\}$ is generated by applying on $\rho$ a set of rank-1 projection operators $\{\ket{z}\bra{z}_B\}$. Consider the spectral decomposition of $\rho_0$, 
\begin{equation*}
	\rho_0=\sum_l\lambda_l\ket{l}\bra{l}\,.
\end{equation*}
where $\lambda_l\in[0,1]$ with $\sum_l\lambda_l=1$. For notational convenience, we define 
\begin{equation}
	\ket{\tilde{\chi}_{z,l}(U)}:= (I_A\otimes \bra{z}_B) \ U\ket{l}\,,
\end{equation}
which enables us to define
\begin{equation}
	\begin{split}
		\mathcal{R}_z(U)&:=\sum_l\lambda_l\ket{\tilde{\chi}_{z,l}(U)}\bra{\tilde{\chi}_{z,l}(U)}\,. \\
		\mathcal{P}_z(U)&:=\text{Tr}[\mathcal{R}_z(U)]\,, 
	\end{split}
\end{equation}
such that $p(z_B)=\mathcal{P}_z(U)$ and $\rho_A(z_B)=\mathcal{R}_z(U)/\mathcal{P}_z(U)$.
Consider the unitary $U_A\in \mathcal{H}_A$, where $\mathcal{H}_A$ is the $d_A$-dimensional Hilbert space over $A$. Let $\Tilde{U}:=U_A\otimes I_B$. We then have,
\begin{equation}
	\begin{split}
		\mathcal{R}_z(\Tilde{U}U)&=U_A\mathcal{R}_z(U)U_A^\dagger\,, \\
		\mathcal{P}_z(\Tilde{U}U)&=\mathcal{P}_z(U)\,.
	\end{split}
\end{equation}
For $U_A\sim\text{Haar}(d_A)$ and $U\sim\text{Haar}(d)$, $\mathcal{R}_z(\Tilde{U}U)$ is equivalent to $\mathcal{R}_z(U)$ in a statistical sense. To compute the moments $\rho^{(k)}_{\eref[\rho_0]}$, we have 
\begin{equation}
	\rho^{(k)}_{\eref[\rho_0]}=\sum_{z_B}p(z_B)\rho_A(z_B)^{\otimes k}=\sum_z \frac{\left(\mathcal{R}_z(U)\right)^{\otimes k}}{\mathcal{P}_z(U)^{k-1}}\,.
	\label{eq:moments}
\end{equation}
We derive the moments of the ensembles by averaging over $U\sim\text{Haar}(d)$. With appropriate concentration inequalities derived from Levy's lemma, it can be shown that a single random unitary can yield this ensemble (see Sec.~II). Using the statistical equivalence of $\mathcal{R}_z(\Tilde{U}U)$ and $\mathcal{R}_z(U)$, we have 
\begin{equation}
	\begin{split}
		\mathbb{E}_{U\sim \text{Haar}(d)}\left[\frac{\left(\mathcal{R}_z(U)\right)^{\otimes k}}{\mathcal{P}_z(U)^{k-1}}\right]&=\mathbb{E}_{U\sim \text{Haar}(d)} \  \mathbb{E}_{U_A \sim \text{Haar}(d_A)} \left[\frac{\left(\mathcal{R}_z(\Tilde{U}U)\right)^{\otimes k}}{\mathcal{P}_z(\Tilde{U}U)^{k-1}}\right]\\
		&=\mathbb{E}_{U\sim \text{Haar}(d)} \  \mathbb{E}_{U_A \sim \text{Haar}(d_A)}\left[\frac{\left(U_A\mathcal{R}_z(U)U_A^\dagger\right)^{\otimes k}}{\mathcal{P}_z(U)^{k-1}}\right]\\
		&=\mathbb{E}_{U\sim \text{Haar}(d)}\left[\frac{1}{\mathcal{P}_z(U)^{k-1}}\right]\sum_{l_1...l_k}\lambda_{l_1}...\lambda_{l_k} \times \\
		&\hspace{0.45cm} \mathbb{E}_{ V \sim \text{Haar}(d)}\left[ \ket{\tilde{\chi}^A_{z,l_1}(V)}\bra{\tilde{\chi}^A_{z,l_1}(V)}\otimes ... \otimes \ket{\tilde{\chi}^A_{z,l_k}(V)}\bra{\tilde{\chi}^A_{z,l_k}(V)}\right]\,,
		\label{eq:Haar-moments}
	\end{split}
\end{equation}
To pass from the second line to the third, we used the fact that $U_A\ket{\tilde{\chi}_{z,l}(U)}\bra{\tilde{\chi}_{z,l}(U)}U_A^\dagger$ is independent of $U$ in the limit $|B|\rightarrow\infty$ (see the independence argument in the latter part of this section for a rigorous justification) and that $U_A\ket{\tilde{\chi}_{z,l}(U)}$ is identically distributed to $\ket{\tilde{\chi}^A_{z,l}(V)}$, where $V\sim\text{Haar}(d)$. 
\\\\
We need not explicitly compute the Haar integral $\mathbb{E}_{U\sim \text{Haar}(d)}\left[\frac{1}{\left[\mathcal{P}_z(U)\right]^{k-1}}\right]$ (and doing so would require further approximations). Rather we absorb all factors in the normalisation $\mathcal{N}$, which is constrained by the fact that the trace of the above tensor is $\mathbb{E}_{U\sim \text{Haar}(d)}\mathcal{P}_z(U)$. This leaves us with
\begin{equation}
	\begin{split}
\rho^{(k)}_{\eref[\rho_0]}=\mathcal{N}\sum_{l_1...l_k}\lambda_{l_1}...\lambda_{l_k} \ \mathbb{E}_{ V \sim \text{Haar}(d)}\left[ \ket{\tilde{\chi}^A_{z,l_1}(V)}\bra{\tilde{\chi}^A_{z,l_1}(V)}\otimes ... \otimes \ket{\tilde{\chi}^A_{z,l_k}(V)}\bra{\tilde{\chi}^A_{z,l_k}(V)}\right]\,.
	\end{split}
\end{equation}
The right part of the above term can be easily evaluated using Weingarten calculus, yielding the result that the projected ensemble is a non-universal ensemble that is a highly non-trivial function of the the eigenvalues of the initial mixed state. The Haar-averaged tensor in Eq.~\ref{eq:moments} can now be represented as
\begin{equation}
	\begin{split}
		\rho^{(k)}_{\eref[\rho_0]}=\mathcal{N}\sum_{\sigma,\tau \in S_k} \ \sum_{l_1...l_{|\tau|}}\lambda_{l_1}^{n_1}...\lambda_{l_{|\tau|}}^{n_{|\tau|}} \ \text{Wg}(\sigma\tau^{-1},d) \  \text{Perm}_{\mathcal{H}_A^{\otimes k}}(\sigma)\,,
	\end{split}
\end{equation}
where $|\tau|$ is the number of cycles and $n_m$ is the order of the $m^{th}$ cycle of $\tau$ with $\sum_{m=1}^{|\tau|}n_m=k$. Since $d\rightarrow\infty$, we have $\text{Wg}(\sigma\tau^{-1},d)\sim \delta_{\sigma,\tau}$, which gives 
\begin{equation}
	\rho^{(k)}_{\eref[\rho_0]}=\mathcal{N}\sum_{\sigma \in S_k} \ \sum_{l_1...l_{|\sigma|}}\lambda_{l_1}^{n_1}...\lambda_{l_{|\sigma|}}^{n_{|\sigma|}} \  \text{Perm}_{\mathcal{H}_A^{\otimes k}}(\sigma)\,.
	\label{eq:moments-calc}
\end{equation}
Since the above term is independent of $z$, the summation of these terms over $z$ merely introduces a constant factor, which is absorbed in the normalization $\mathcal{N}$. The normalization is determined by the fact that the trace is unity. Defining 
\begin{equation}
h_\sigma(\rho_0)=\text{Tr}\rho_0^{n_1}...\text{Tr}\rho_0^{n_{|\sigma|}}=\sum_{l_1...l_{|\sigma|}}\lambda_{l_1}^{n_1}...\lambda_{l_{|\sigma|}}^{n_{|\sigma|}}\,,    
\end{equation}
where $|\sigma|$ is the number of cycles and $n_m$ is the order of the $m^{th}$ cycle of $\sigma$ with $\sum_{m=1}^{|\sigma|}n_m=k$, we obtain Eq.~\ref{eq:eref-moments} and Eq.~\ref{eq:hsigma}. 

\subsection{The independence argument}
From the translation invariance of the Haar measure, it is evident that the distribution of $U_A\ket{\chi^A_{z,l}(U)}\bra{\chi^A_{z,l}(U)}U_A^\dagger$ is independent of the distribution of $\ket{\chi^A_{z,l}(U)}\bra{\chi^A_{z,l}(U)}$, where
$\ket{\chi^A_{z,l}(U)}:= \frac{\ket{\tilde{\chi}_{z,l}(U)}}{\sqrt{\langle\tilde{\chi}_{z,l}(U)|\tilde{\chi}_{z,l}(U)\rangle}}$ is a normalised state. For $d\gg 1$, we show that 
\begin{equation}
	\ket{\tilde{\chi}_{z,l}(U)}\bra{\tilde{\chi}_{z,l}(U)}\sim \tfrac{1}{d_B}\ket{\chi^A_{z,l}(U)}\bra{\chi^A_{z,l}(U)}    
	\label{eq:approx-norm}
\end{equation}
where $d_B=d/d_A$. This justifies that $U_A\ket{\tilde{\chi}_{z,l}}\bra{\tilde{\chi}_{z,l}}U_A^\dagger$ is independent of $U$ in the limit $d\rightarrow\infty$.
To show Eq.~\ref{eq:approx-norm}, we prove that the distribution of $\langle\tilde{\chi}_{z,l}|\tilde{\chi}_{z,l}\rangle$ is sharply peaked around $1/d_B$ for $d\gg 1$ using Levy's lemma. 
\\\\
Consider a Haar random state $\ket{\Phi}$, where $\ket{\Phi}\sim\text{Haar}(d)$. We define $\ket{\Phi_z}:=(I_A\otimes \bra{z}_B) \ket{\Phi}$ and $G(\ket{\Phi}):=\langle \Phi_z|\Phi_z\rangle$. Note that $G(\ket{\Phi})=\langle\tilde{\chi}_{z,l}|\tilde{\chi}_{z,l}\rangle$. A quick calculation yields $\mathbb{E}_{\ket{\Phi}\sim\text{Haar}(d)}G(\ket{\Phi})=d^{-1}_B$. We now state Levy's lemma. 
\\\\
\textbf{Lemma 1.} (Levy's lemma for random states, see~\cite{cotler2023emergent}) \textit{Let Let $f:\mathbb{S}^{2d-1}\rightarrow\mathbb{R}$ be an $\eta$-Lipschitz function. Then for any $\delta\geq 0$, we have}
\begin{equation}    \text{Prob}_{\ket{\Phi}\sim\text{Haar}(d)}[|f(\ket{\Phi})-\mathbb{E}_{\ket{\Psi}\sim\text{Haar}(d)}[f(\ket{\Psi})]|\geq \delta]\leq 2\exp\left(-\frac{2d\delta^2}{9\pi^3\eta^2}\right)
\end{equation}
Borrowing the notational framework from~\cite{cotler2023emergent}, given $\Vec{v}\in \mathbb{S}^{2d-1}$, we have $\ket{\Phi}=\begin{bmatrix}
	I_d & iI_d\\
\end{bmatrix}.~\Vec{v}$, where $\ket{\Phi}$ is a normalised state in $\mathbb{C}^d$ and $I_d$ is a $d\times d$ identity matrix. This allows us to represent the arguments of $f$ either as normalised states in $\mathbb{C}^d$ or real vectors in $\mathbb{S}^{2d-1}$. Here, $\Vec{v}$ is represented as $\begin{bmatrix}
	\Vec{a}\\ \Vec{b}
\end{bmatrix}$, where $\Vec{a},\Vec{b}\in \mathbb{R}^d$ and $\Vec{a}.\Vec{a}+\Vec{b}.\Vec{b}=1$. To apply Levy's lemma to $G(\ket{\Phi})$, we prove the following lemma:
\\\\
\textbf{Lemma 2.} A Lipschitz constant for $G$ is $\eta=2$.
\\\\
\textit{Proof.} Since $G$ is a differentiable function, we can take any $\eta$ such that 
\begin{equation*}
	\eta\geq\left\Vert\frac{d}{d\Vec{v}}G(\ket{\Phi})\right\Vert_2    
\end{equation*}
We have 
\begin{equation*}
	\begin{split}
		\left\Vert\frac{d}{d\Vec{v}}G(\ket{\Phi}\right\Vert_2 &=  \left\Vert\frac{d}{d\Vec{v}}\left(\Vec{v}^\text{T}.\begin{bmatrix}
			\mathbb{I} \\ -i\mathbb{I}
		\end{bmatrix} . \left(I_A\otimes\ket{z}\bra{z}_B\right).\begin{bmatrix}
			\mathbb{I} & i\mathbb{I}
		\end{bmatrix}. ~\Vec{v}\right) 
		\right\Vert_2 \\
		&=\left\Vert\begin{bmatrix}
			2P_z & \mathbb{O} \\
			\mathbb{O} & 2P_z
		\end{bmatrix}.~\Vec{v}\right\Vert_2 \\
		&=2\left(\Vec{a}.P_z.\Vec{a}+\Vec{b}.P_z.\Vec{b}\right) \leq 2
	\end{split}
\end{equation*}
where $P_z=\mathbb{I}_A\otimes\ket{z}\bra{z}_B$ and $\mathbb{O}$ and $\mathbb{I}$ are $d\times d$ null and identity matrices, respectively. Combining this result with Lemma 1 gives us the concentration inequality: 
\begin{equation}    
	\text{Prob}_{\ket{\Phi}\sim\text{Haar}(d)}\left[\Big|G(\ket{\Phi})-\frac{1}{d_B}\Big|\geq \delta\right]\leq 2\exp\left(-\frac{d\delta^2}{18\pi^3}\right)
	\label{eq:conc-ineq}
\end{equation}
which concludes the justification for the independence argument.
\section{Typicality of $\eref[\rho_0]$ over random unitaries}
\label{sec:typicality}
In this section, we provide evidence for the argument that the moments of $\eref[\rho_0]$ computed by averaging over $U\sim \text{Haar}(d)$ is typical of $\text{Haar}(d)$ for large $d$. To this end, we demonstrate that the distribution of each component of the moments of the projected of the ensemble in some basis is strongly concentrated about its Haar-averaged value for $d\gg 1$. The key to this argument is Levy's lemma for a random unitary, stated as follows: \\\\
\textbf{Lemma 3.} (Levy's lemma for a random unitary, see~\cite{ledoux2001concentration}) \textit{Let f be an $\eta$-Lipschitz function on $U(d)$. Then for any $\delta\geq 0$, we have}
\begin{equation}
	\text{Prob}_{U\sim\text{Haar(d)}}\left[|f(U)-\mathbb{E}_{V\sim\text{Haar}(d)}\left[f(V)\right]|\geq \delta\right]\leq 4\exp\left(-\frac{2d\delta^2}{9\pi^3\eta^2}\right).
\end{equation} 

We will now define the function of interest to apply this lemma to the projected ensemble of the form in [eq (1)]. Following the notations from Appx.~F of~\cite{cotler2023emergent}, we define $\ket{i}=\otimes_{k=1}^t\ket{i^{(k)}}$ and $\ket{j}=\otimes_{k=1}^t\ket{j^{(k)}}$, where each $\ket{i^{(k)}},\ket{j^{(k)}}\in\mathcal{H}_A$. Consider the function $f_{ij}:U(d)\rightarrow \mathbb{R}$ defined by 
\begin{equation}
	\begin{split}
		f_{ij}(U)&=\sum_{z_B}\bra{i}\left(\frac{\left(\sum_m\lambda_m\ket{\tilde{\chi}_{z,m}}\bra{\tilde{\chi}_{z,m}}\right)^{\otimes k}}{\left(\sum_m\lambda_m\braket{\tilde{\chi}_{z,m}|\tilde{\chi}_{z,m}}\right)^{k-1}}\right)\ket{j}\\
		&=\sum_{z_B}\frac{\prod_{l=1}^k\sum_{m_l}\lambda_{m_l}\braket{i^{(l)}|\tilde{\chi}_{z,m}}\braket{\tilde{\chi}_{z,m}|j^{(l)}}}{\left(\sum_m\lambda_m\braket{\tilde{\chi}_{z,m}|\tilde{\chi}_{z,m}}\right)^{k-1}}\,,
	\end{split}
\end{equation}
where $\ket{\tilde{\chi}_{z,m}}= \left(\mathbb{I}_{D_A}\otimes\bra{z_B}\right)U\ket{m}$ (defined in Sec.~I) encodes the $U$ dependence. We now prove the following lemma: \\\\
\textbf{Lemma 4.} A Lipschitz constant for $f_{ij}$ is $\eta=2(2k-1)$. \\\\

\textit{Proof.} Any $\eta$ such that
\begin{equation*}
	\eta\geq \left\Vert\frac{d}{dU}f_{ij}(U)\right\Vert_2\,,
\end{equation*}
can be a Lipschitz constant since $f_{ij}$ is differentiable. To obtain this bound, we express the above matrix derivative as an array of vector derivatives in a particular basis. To that end, we define $\vec{u}_m\in\mathbb{S}^{2d-1}$ such that $U\ket{m}=\imat.\vec{u}_m$. We express $\frac{d}{dU}f_{ij}(U)$ as $\left[\frac{d}{d\vec{u}_m}f_{ij}(U)\right]_m$, where $\mathbb{I}$ denotes the $d\times d$ identity matrix. We have 
\begin{equation}
	\begin{split}
		\frac{d}{d\vec{u}_m}f_{ij}(U)=&\sum_{l'=1}^k\sum_z\frac{\prod_{l\neq l'}\sum_{m_l}\lambda_{m_l}\braket{i^{(l)}|\tilde{\chi}_{z,m_l}}\braket{\tilde{\chi}_{z,m_l}|j^{(l)}}}{\left(\sum_m\lambda_m\braket{\tilde{\chi}_{z,m}|\tilde{\chi}_{z,m}}\right)^{k-1}}\sum_{m_{l'}}\lambda_{m_{l'}}\frac{d}{d\vec{u}_m}\braket{i^{(l')}|\tilde{\chi}_{z,m_{l'}}}\braket{\tilde{\chi}_{z,m_{l'}}|j^{(l')}}\\
		&-(k-1)\sum_z\frac{\prod^k_{l=1}\sum_{m_l}\lambda_{m_l}\braket{i^{(l)}|\tilde{\chi}_{z,m_l}}\braket{\tilde{\chi}_{z,m_l}|j^{(l)}}}{\left(\sum_m\lambda_m\braket{\tilde{\chi}_{z,m}|\tilde{\chi}_{z,m}}\right)^{k}}\sum_{n}\lambda_{n}\frac{d}{d\vec{u}_m}\braket{\tilde{\chi}_{z,n}|\tilde{\chi}_{z,n}}\\
		=&\sum_{l'=1}^k\sum_z\frac{\prod_{l\neq l'}\sum_{m_l}\lambda_{m_l}\braket{i^{(l)}|\tilde{\chi}_{z,m_l}}\braket{\tilde{\chi}_{z,m_l}|j^{(l)}}}{\left(\sum_m\lambda_m\braket{\tilde{\chi}_{z,m}|\tilde{\chi}_{z,m}}\right)^{k-1}}\left\{\begin{bmatrix}
			M^+_{l'}& -iM^-_{l'}\\
			iM^+_{l'} & M^+_{l'}
		\end{bmatrix}\otimes\ket{z}\bra{z}_B\right\}.\vec{u}_m\lambda_{m}\\
		&-(k-1)\sum_z\frac{\prod^k_{l=1}\sum_{m_l}\lambda_{m_l}\braket{i^{(l)}|\tilde{\chi}_{z,m_l}}\braket{\tilde{\chi}_{z,m_l}|j^{(l)}}}{\left(\sum_m\lambda_m\braket{\tilde{\chi}_{z,m}|\tilde{\chi}_{z,m}}\right)^{k}}\begin{bmatrix}
			P_z & 0 \\
			0 & P_z
		\end{bmatrix}.\vec{u}_m\times 2\lambda_m\,,
	\end{split}
\end{equation}
where we have defined $M^{\pm}_{l'}=\ket{i^{(l')}}\bra{j^{(l')}}\pm\ket{j^{(l')}}\bra{i^{(l')}}$ and have explicitly evaluated the derivatives. This gives
\begin{equation}
	\begin{split}
			\frac{d}{dU}f_{ij}(U)=&\sum_{l'=1}^k\sum_z\frac{\prod_{l\neq l'}\sum_{m_l}\lambda_{m_l}\braket{i^{(l)}|\tilde{\chi}_{z,m_l}}\braket{\tilde{\chi}_{z,m_l}|j^{(l)}}}{\left(\sum_m\lambda_m\braket{\tilde{\chi}_{z,m}|\tilde{\chi}_{z,m}}\right)^{k-1}}\left\{\begin{bmatrix}
			M^+_{l'}& -iM^-_{l'}\\
			iM^+_{l'} & M^+_{l'}
		\end{bmatrix}\otimes\ket{z}\bra{z}_B\right\}.VD\\
		&-(k-1)\sum_z\frac{\prod^k_{l=1}\sum_{m_l}\lambda_{m_l}\braket{i^{(l)}|\tilde{\chi}_{z,m_l}}\braket{\tilde{\chi}_{z,m_l}|j^{(l)}}}{\left(\sum_m\lambda_m\braket{\tilde{\chi}_{z,m}|\tilde{\chi}_{z,m}}\right)^{k}}\begin{bmatrix}
			P_z & 0 \\
			0 & P_z
		\end{bmatrix}.VD\times 2\,,
	\end{split}
\end{equation}
where $V:\mathbb{C}^d\rightarrow\mathbb{S}^{2d-1}$ is an matrix such that $U=\imat.V$ and $D=\text{dia}\left(\lambda_1,\lambda_2...\lambda_m\right)$. Taking the norm and applying the triangle inequality gives
\begin{equation}
	\begin{split}
		\left\Vert\frac{d}{dU}f_{ij}(U)\right\Vert_2\leq~&\Bigg\Vert\sum_{l'=1}^k\sum_z\frac{\prod_{l\neq l'}\sum_{m_l}\lambda_{m_l}\braket{i^{(l)}|\tilde{\chi}_{z,m_l}}\braket{\tilde{\chi}_{z,m_l}|j^{(l)}}}{\left(\sum_m\lambda_m\braket{\tilde{\chi}_{z,m}|\tilde{\chi}_{z,m}}\right)^{k-1}}\left\{\begin{bmatrix}
			M^+_{l'}& -iM^-_{l'}\\
			iM^+_{l'} & M^+_{l'}
		\end{bmatrix}\otimes\ket{z}\bra{z}_B\right\}.VD\Bigg\Vert_2\\
		&+2(k-1)\Bigg\Vert\sum_z\frac{\prod^k_{l=1}\sum_{m_l}\lambda_{m_l}\braket{i^{(l)}|\tilde{\chi}_{z,m_l}}\braket{\tilde{\chi}_{z,m_l}|j^{(l)}}}{\left(\sum_m\lambda_m\braket{\tilde{\chi}_{z,m}|\tilde{\chi}_{z,m}}\right)^{k}}\begin{bmatrix}
			P_z & 0 \\
			0 & P_z
		\end{bmatrix}.VD\Bigg\Vert_2\,,
	\label{eq:F8}
	\end{split}
\end{equation}
We now individually bound each of the terms on the RHS. For the first term, writing $b_{l',z}=\left(\prod_{l\neq l'}\sum_{m_l}\lambda_{m_l}\braket{i^{(l)}|\tilde{\chi}_{z,m_l}}\braket{\tilde{\chi}_{z,m_l}|j^{(l)}}\right)/\left(\sum_m\lambda_m\braket{\tilde{\chi}_{z,m}|\tilde{\chi}_{z,m}}\right)^{k-1}$ and $M_{l'}=\begin{bmatrix}
	M^+_{l'}& -iM^-_{l'}\\
	iM^+_{l'} & M^+_{l'}
\end{bmatrix}$, we have
\begin{equation}
	\begin{split}
		\Bigg\Vert\sum_{l'=1}^k\sum_z b_{l',z}\left\{M_{l'}\otimes\ket{z}\bra{z}_B\right\}VD\Bigg\Vert_2 =& \left(\sum_{l',p'=1}^k\sum_{z}b^*_{p',z}b_{l',z}\text{Tr}\left[DV^\dagger\left(M^\dagger_{p'}M_{l'}\otimes\ket{z}\bra{z}_B\right)VD\right]\right)^{1/2}\\
		\leq & \left(\sum_{l',p'=1}^k\sum_{z}|b^*_{p',z}||b_{l',z}|\text{Tr}\left[DV^\dagger \left(|M^\dagger_{p'}M_{l'}|\otimes\ket{z}\bra{z}_B\right) VD\right]\right)^{1/2},
	\end{split}
\label{eq:F10}
\end{equation}
where $|A|:=\sqrt{A^\dagger A}$. We note that
\begin{equation}
	\begin{split}
		|b_{l',z}|=&\left\vert\text{Tr}\left\{\left(\otimes_{l\neq l'}\ket{j^{(l)}}\bra{i^{(l)}}\right).\left(\frac{\left(\sum_m\lambda_m\ket{\tilde{\chi}_{z,m}}\bra{\tilde{\chi}_{z,m}}\right)^{\otimes k-1}}{\left(\sum_m\lambda_m\braket{\tilde{\chi}_{z,m}|\tilde{\chi}_{z,m}}\right)^{k-1}}\right)\right\}\right\vert\\
		\leq& \left\Vert \otimes_{l\neq l'}\ket{j^{(l)}}\bra{i^{(l)}}\right\Vert_2 \left\Vert\frac{\sum_m\lambda_m\ket{\tilde{\chi}_{z,m}}\bra{\tilde{\chi}_{z,m}}}{\sum_m\lambda_m\braket{\tilde{\chi}_{z,m}|\tilde{\chi}_{z,m}}}\right\Vert_2^{k-1}\leq 1\,,
	\end{split}
\end{equation}
since $	\left\Vert \otimes_{l\neq l'}\ket{j^{(l)}}\bra{i^{(l)}}\right\Vert_2=1$ and
\begin{equation}
	\begin{split}
		\left\Vert\frac{\sum_m\lambda_m\ket{\tilde{\chi}_{z,m}}\bra{\tilde{\chi}_{z,m}}}{\sum_m\lambda_m\braket{\tilde{\chi}_{z,m}|\tilde{\chi}_{z,m}}}\right\Vert_2=\left(\frac{\sum_{m,m'}\lambda_m\lambda_{m'}|\braket{\tilde{\chi}_{z,m}|\tilde{\chi}_{z,m'}}|^2}{\sum_{m,m'}\lambda_m\lambda_{m'}\braket{\tilde{\chi}_{z,m}|\tilde{\chi}_{z,m}}\braket{\tilde{\chi}_{z,m'}|\tilde{\chi}_{z,m'}}}\right)&\leq 1\,,		
	\end{split}
\end{equation}
where the inequality follows from the Cauchy-Schwarz inequality. Eq.~\ref{eq:F10} is less than or equal to
\begin{equation}
	\begin{split}
		\left(\sum_{l',p'=1}^k\sum_{z}\text{Tr}\left[DV^\dagger \left(|M^\dagger_{p'}M_{l'}|\otimes\ket{z}\bra{z}_B\right) VD\right]\right)^{1/2}&=	\left(\sum_{l',p'=1}^k\text{Tr}\left[DV^\dagger \left(|M^\dagger_{p'}M_{l'}|\otimes\mathbb{I}_B\right) VD\right]\right)^{1/2}\\
		&=\left(\sum_{l',p'=1}^k\sum_m \lambda_m^2 \vec{u}^T_m. \left(|M^\dagger_{p'}M_{l'}|\otimes\mathbb{I}_B\right) .\vec{u}_m \right)^{1/2}\\
		&\leq \left(\sum_{l',p'=1}^k\Vert M^\dagger_{p'}M_{l'}\Vert_\infty\right)^{1/2},
	\end{split}
\end{equation}
where we have used the monotonicity of norms and that $\lambda_m\in[0,1]$ with $\sum_m\lambda_m=1$, in addition to $\vec{u}_m^T.\vec{u}_m=1$. Since $\Vert M_{p'}^\dagger M_{l'}\Vert_\infty\leq \Vert M_{p'}^\dagger\Vert_\infty \Vert M_{l'}\Vert_\infty\leq 4$, the above is $\leq 2k$.
\\\\
Now we bound the second term in the RHS of Eq.~\ref{eq:F8}.\linebreak Letting $c_z=\left(\prod^k_{l=1}\sum_{m_l}\lambda_{m_l}\braket{i^{(l)}|\tilde{\chi}_{z,m_l}}\braket{\tilde{\chi}_{z,m_l}|j^{(l)}}\right)/\left(\sum_m\lambda_m\braket{\tilde{\chi}_{z,m}|\tilde{\chi}_{z,m}}\right)^{k}$, the term can be written as
\begin{equation}
	2(k-1)\left(\sum_z|c_z|^2\text{Tr}\left[DV^\dagger.\begin{bmatrix}
		P_z & 0 \\
		0 & P_z
	\end{bmatrix}.VD\right]\right)^{1/2}=	2(k-1)\left(\sum_z|c_z|^2\sum_m`\lambda_m^2\vec{u}_m^T.\begin{bmatrix}
	P_z & 0 \\
	0 & P_z
\end{bmatrix}.\vec{u}_m\right)^{1/2}.
\label{eq:F12}
\end{equation}
Using the bound 
\begin{equation}
	\begin{split}
		|c_z|=&\left\vert\text{Tr}\left\{\left(\otimes_{l\neq l'}\ket{j^{(l)}}\bra{i^{(l)}}\right).\left(\frac{\left(\sum_m\lambda_m\ket{\tilde{\chi}_{z,m}}\bra{\tilde{\chi}_{z,m}}\right)^{\otimes k}}{\left(\sum_m\lambda_m\braket{\tilde{\chi}_{z,m}|\tilde{\chi}_{z,m}}\right)^{k}}\right)\right\}\right\vert\\
		\leq& \left\Vert \otimes_{l\neq l'}\ket{j^{(l)}}\bra{i^{(l)}}\right\Vert_2 \left\Vert\frac{\sum_m\lambda_m\ket{\tilde{\chi}_{z,m}}\bra{\tilde{\chi}_{z,m}}}{\sum_m\lambda_m\braket{\tilde{\chi}_{z,m}|\tilde{\chi}_{z,m}}}\right\Vert_2^{k}\leq 1\,,
	\end{split}
\end{equation}
we can upper bound Eq.~\ref{eq:F12} by 
\begin{equation}
	\begin{split}
		2(k-1)\left(\sum_z\sum_m`\lambda_m^2\vec{u}_m^T.\begin{bmatrix}
			P_z & 0 \\
			0 & P_z
		\end{bmatrix}.\vec{u}_m\right)^{1/2}&=2(k-1)\left(\sum_z\sum_m`\lambda_m^2\vec{u}_m^T.\left(\begin{bmatrix}
		\mathbb{I}_A & 0 \\
		0 & \mathbb{I}_A
	\end{bmatrix}\otimes P_z\right).\vec{u}_m\right)^{1/2}\\
&\leq 2(k-1)\,.
	\end{split}
\end{equation}
Putting our bounds together, we have 
\begin{equation}
	\left\Vert\frac{d}{dU}f_{ij}(U)\right\Vert_2\leq 2k+2(k-1)=2(2k-1)\,.
\end{equation}
which concludes our proof of Lemma 4. Combining this result with Lemma 3, we obtain the concentration inequality:
\begin{equation}
	\text{Prob}_{U\sim\text{Haar(d)}}\left[|f_{ij}(U)-\mathbb{E}_{V\sim\text{Haar}(d)}\left[f_{ij}(V)\right]|\geq \delta\right]\leq 4\exp\left(-\frac{d\delta^2}{18\pi^3(2k-1)^2}\right),
\end{equation}
which explicitly shows typicality for $d\gg 1$.
\section{Connection to the Scrooge Ensemble}

In this section, we prove the relation in Eq.~\ref{eq:eref-scrooge-res}, which demonstrates that $\eref[\rho_0]$ can be obtained by partially tracing out states from an appropriately defined Scrooge ensemble. We first lay out a few definitions for notational convenience. The computational basis of the Hilbert space of $A\cup B$ (or of the Hilbert space of $X$) is given by $\{\ket{l}\}_{l=1}^{D_X}$. We now define the kets
\begin{equation}
    \begin{split}
    \ket{\tilde{\chi}_{z,l}}:&= \left(\mathbb{I}_{D_A}\otimes\bra{z_B}\right)U\ket{l}\\
\ket{\tilde{\psi}_z}:&=\sum_l\ket{l}\otimes \ket{\tilde{\chi}_{z,l}}\,,
    \end{split}
\end{equation}
where $U\sim\text{Haar}(D_{AB})$. This enables a compact representation of the states
 $\ket{\phi_{XA}(z_B)}$ and the probabilities $p(z_B)$, such that
 
 \begin{equation}
     \begin{split}
          \ket{\phi_{XA}(z_B)}&=\frac{\sqrt{\rho_{XA}}\ket{\tilde{\psi}_z}}{\sqrt{\bra{\tilde{\psi}_z}\rho_{XA}\ket{\tilde{\psi}_z}}}\\\\
 p(z_B)&=D_A\bra{\tilde{\psi}_z}\rho_{XA}\ket{\tilde{\psi}_z}\,,
     \end{split}
 \end{equation}
 
where $\rho_{XA} = (\rho_0^T \otimes \mathbb{I}_{D_A})/D_A$. The moments of $\eref[\rho_0]$ are given by
\begin{equation}
	\begin{split}
\rho^{(k)}_{\eref[\rho_0]}&=\sum_{z_B}p(z_B)\left[{\rm Tr}_X\ket{\phi_{XA}(z_B)}\bra{\phi_{XA}(z_B)}\right]^{\otimes k}\\
&=\sum_{z_B}\int dU_{\text{Haar}(D_{AB})}D_A\frac{\left(\text{Tr}_X\left[\sqrt{\rho_{XA}}\ket{\tilde{\psi}_z}\bra{\tilde{\psi}_z}\sqrt{\rho_{XA}}\right]\right)^{\otimes k}}{\bra{\tilde{\psi}_z}\rho_{XA}\ket{\tilde{\psi}_z}^{k-1}}\,,
\label{eq:moments-step1}
\end{split}
\end{equation}
where the Haar integral appears due to typicality in the limit $D_B\rightarrow\infty$. We note that the distribution of $\braket{\tilde{\psi}_z|\tilde{\psi}_z}$ is highly concentrated around $D_A$ in the limit $D_B\rightarrow\infty$ (this follows from the concentration of $\braket{\tilde{\chi}_{z,l}|\tilde{\chi}_{z,l}}$, which was proven in Sec.~I). This allows us to plug in the normalised state $\ket{\psi_z}=\tfrac{\ket{\tilde{\psi}_z}}{\sqrt{\braket{\tilde{\psi}_z|\tilde{\psi}_z}}}$ into Eq.~\ref{eq:moments-step1}, giving
\begin{equation}
\rho^{(k)}_{\eref[\rho_0]}=\sum_{z_B}\int dU_{\text{Haar}(D_{AB})}D_A^2\frac{\left(\text{Tr}_X\left[\sqrt{\rho_{XA}}\ket{\psi_z}\bra{\psi_z}\sqrt{\rho_{XA}}\right]\right)^{\otimes k}}{\bra{\psi_z}\rho_{XA}\ket{\psi_z}^{k-1}} \,.
\label{eq:moments-step2}
\end{equation}

Now, we prove that the continuous set $\mathcal{E}_U=\{dU_{\text{Haar}(D_{AB})},\ket{\psi_z}\}$ forms an exact state $k$-design (for a fixed $k$) over $\mathbb{C}^{D_{XA}}$ in the limit $D_B\rightarrow\infty$. The moments of $\mathcal{E}_U$ are given by
\begin{equation}
    \begin{split}
        \rho^{(k)}_{\mathcal{E}_U}&=\int dU_{\text{Haar}(D_{AB})}\left(\ket{\psi_z}\bra{\psi_z}\right)^{\otimes k}\\
        &=\int dU_{\text{Haar}(D_{AB})} \frac{\left(\sum_{l,l'}\ket{l'}\bra{l}\otimes \ket{\tilde{\chi}_{z,l'}}\bra{\tilde{\chi}_{z,l}}\right)^{\otimes k}}{\left(\sum_l\braket{\tilde{\chi}_{z,l}|\tilde{\chi}_{z,l}}\right)^k}\,.
        \label{eq:moments-step3}
    \end{split}
\end{equation}
The above integral is invariant under the transformation $U\rightarrow \left(U_A\otimes\mathbb{I}_{D_B}\right) U$, where $U_A\sim\text{Haar}(D_A)$ is statistically independent of $U$. Plugging this into Eq.~\ref{eq:moments-step3}

\begin{equation}
    \begin{split}
    \rho^{(k)}_{\mathcal{E}_U}&=\int dU_{A,\text{Haar}(D_A)} \int dU_{\text{Haar}(D_{AB})} \frac{\left(\sum_{l,l'}\ket{l'}\bra{l}\otimes \left(U_A\ket{\tilde{\chi}_{z,l'}}\bra{\tilde{\chi}_{z,l}}U_A^\dagger\right)\right)^{\otimes k}}{\left(\sum_l\braket{\tilde{\chi}_{z,l}|\tilde{\chi}_{z,l}}\right)^k}\,.
    \label{eq:moments-step4}
    \end{split}
\end{equation}

The above integral factorises since $U_A\ket{\tilde{\chi}_{z,l}}\bra{\tilde{\chi}_{z,l}}U_A^\dagger$ is statistically independent of $U$ in the limit of $D_B\rightarrow\infty$ (see proof in Sec.~I). In addition, $U_A\ket{\tilde{\chi}_{z,l}}$ is identically distributed to $\ket{\tilde{\zeta}_{z,l}}:=\left(\mathbb{I}_{D_A}\otimes\bra{z_B}\right)V\ket{l}$, where $V\sim\text{Haar}(D_{AB})$. This gives
\begin{equation}
\begin{split}
\rho^{(k)}_{\mathcal{E}_U}&=\int dV_{\text{Haar}(D_{AB})}\left(\sum_{l,l'}\ket{l'}\bra{l}\otimes\ket{\tilde{\zeta}_{z,l'}}\bra{\tilde{\zeta}_{z,l}}\right)^{\otimes k}\int dU_{\text{Haar}(D_{AB})} \left(\sum_l\braket{\tilde{\chi}_{z,l}|\tilde{\chi}_{z,l}}\right)^{-k}\\
&=\mathcal{N}\sum_{\substack{{l_1...l_k}\\{l_1'...l_k'}}}\left(\ket{l_1'}\bra{l_1}\otimes...\otimes \ket{l_k'}\bra{l_k)}\otimes \sum_{\sigma,\tau\in S_k}\left(\text{Wg}(\sigma\tau^{-1},D_{AB})~\delta_{l'_1,l_{\tau(1)}}...\delta_{l'_k,l_{\tau(k)}} \text{Perm}_{\mathcal{H}_A^{\otimes k}}(\sigma)\right)\right)\\
&=\mathcal{N}\sum_{\sigma\in S_k}\sum_{l_1...l_k}\ket{l_{\sigma(1)}}\bra{l_1}\otimes...\otimes \ket{l_{\sigma(k)}}\bra{l_k}\otimes \text{Perm}_{\mathcal{H}_A^{\otimes k}}(\sigma) \\
&=\frac{\sum_{\sigma\in S_k} \text{Perm}_{(\mathcal{H}_X\otimes\mathcal{H}_A)^{\otimes k}}(\sigma)}{D_{AX}(D_{AX}+1)...(D_{AX}+k-1)}\,,
\label{eq:moments-step5}
\end{split}
\end{equation}
where the intermediate equalities hold modulo an ordering of the tensor products. The integral in the right of the first line of Eq.~\ref{eq:moments-step5} is absorbed by a normalisation constant $\mathcal{N}$, which can be evaluated by taking the trace of the numerator in the last line of Eq.~\ref{eq:moments-step5}. Since $\mathcal{E}_U$ forms an exact state $t$-design over $\mathbb{C}^{D_{XA}}$, the integral $\int_{U\sim \text{Haar}(D_{AB})}$ in Eq.~\ref{eq:moments-step2} can be replaced with the integral $\int d\psi_{\text{Haar}(D_{XA})}$ (see Ref.~\cite{ho2022exact} for a detailed justification) and $\psi_z$ with $\psi$, which gives us

\begin{equation}
    \begin{split}
  \rho^{(k)}_{\eref[\rho_0]}&=\sum_{z_B}\int d\psi_{\text{Haar}(D_{XA})}~D_A^2\frac{\left(\text{Tr}_X\left[\sqrt{\rho_{XA}}\ket{\psi}\bra{\psi}\sqrt{\rho_{XA}}\right]\right)^{\otimes k}}{\bra{\psi}\rho_{XA}\ket{\psi}^{k-1}}\\
  &= \int d\psi_{\text{Haar}(D_{XA})}~D_{XA}\frac{\left(\text{Tr}_X\left[\sqrt{\rho_{XA}}\ket{\psi}\bra{\psi}\sqrt{\rho_{XA}}\right]\right)^{\otimes k}}{\bra{\psi}\rho_{XA}\ket{\psi}^{k-1}}\,,
    \end{split}
\end{equation}
which is Eq.~\ref{eq:eref-scrooge-res} of the main text.

\section{Conditions for $\eref[\rho_0]$ to reduce to the generalized Hilbert-Schmidt ensemble}
 
In this section, we explicitly demonstrate how $\eref[\rho_0]$ reduces to an appropriate gHSe (at the level of their respective moments) under the conditions that were laid out in the main text. The gHSe is defined by tracing out subsystem $R$ from Haar random states on $R \cup A$,
\eq{
\mathcal{E}_\text{gHSe}=\{{\rm Tr}_R[\ket{\psi}\bra{\psi}]:\ket{\psi}\sim {\rm Haar}(D_{RA})\}\,,
\label{eq:gHSe-supplem}
}
The moments of $\eref[\rho_0]$ reduce to those of $\mathcal{E}_\text{gHSe}$ (with $D_R=\text{rank}(\rho_0)$) if the non-zero part of the spectrum of $\rho_0$ is flat. Under this condition, the moments of $\eref[\rho_0]$ acquire a simple form, which we obtain by plugging in $\text{Tr}\rho_0^n=D_R^{1-n}$ into Eq.~\ref{eq:moments-calc},

\begin{equation}
    \begin{split}
\rho^{(k)}_{\eref[\rho_0]}&=\frac{\sum_{\sigma \in S_k} \ D_R^{1-n_1}...~D_R^{1-n_{|\sigma|}} \  \text{Perm}_{\mathcal{H}_A^{\otimes k}}(\sigma)}{\sum_{\sigma \in S_k} \ D_R^{1-n_1}...~D_R^{n_{1-|\sigma|}}D_A^{|\sigma|}}\\
        &=\frac{\sum_{\sigma \in S_k} \ D_R^{|\sigma|}\  \text{Perm}_{\mathcal{H}_A^{\otimes k}}(\sigma)}{\sum_{\sigma \in S_k} \ D_{RA}^{|\sigma|}}\\
        \label{eq:eref_flat}
    \end{split}
\end{equation}
where we used the identity $\sum_i^{|\sigma|}n_i=k$. This exactly matches with the corresponding moment of the gHSe which, according to Eq.~\ref{eq:gHSe-supplem}, is given by, 
\begin{equation}
    \begin{split}
        \rho^{(k)}_{\mathcal{E}_\text{gHSe}}&=\int d\psi_{\text{Haar}(D_{RA})}\left(\text{Tr}_R\left[\ket{\psi}\bra{\psi}\right]\right)^{\otimes k}\\
        &=\frac{\sum_{\sigma \in S_k} \ \text{Tr}\left[\text{Perm}_{\mathcal{H}_R^{\otimes k}}(\sigma)\right]\text{Perm}_{\mathcal{H}_A^{\otimes k}}(\sigma)}{\sum_{\sigma \in S_k} \ D_{RA}^{|\sigma|}}\\
        &=\frac{\sum_{\sigma \in S_k} \ D_R^{|\sigma|}\text{Perm}_{\mathcal{H}_A^{\otimes k}}(\sigma)}{\sum_{\sigma \in S_k} \ D_{RA}^{|\sigma|}}
    \end{split}
\end{equation}
where the intermediate equalities hold modulo an ordering of the tensor products.

\section{Mixed-state deep-thermalisation timescales}
\subsection{Scaling of timescale with $k$}
In the main text, we demonstrated for the kicked Ising chain that in the limit $|B|\to\infty$, the projected ensemble approaches $\eref[\rho_0]$, with $\Delta_k \to 0$ as $t\to\infty$.
It is therefore natural to ask what hierarchy of timescales emerges as a function of $k$ and $|S|$.
Using finite-size scaling analyses (details are provided in the following subsection), we first establish that in the limit $|B|\to\infty$, the dynamical behaviour of $\Delta_k(t)$ with $t$ is independent of $|S|$.
This implies that the mixed-state deep-thermalisation timescales are independent of subsystem size $|S|$.
To extract the dependence of the timescales on $k$, as is customary (see, for example, Ref.~\cite{cotler2023emergent}), we set a fixed threshold $\epsilon$ and determine the timescale $t_k$ at which $\Delta_k(t)$ drops below $\epsilon$.

\begin{figure}[t]
    \centering
    \includegraphics[width=.5\linewidth]{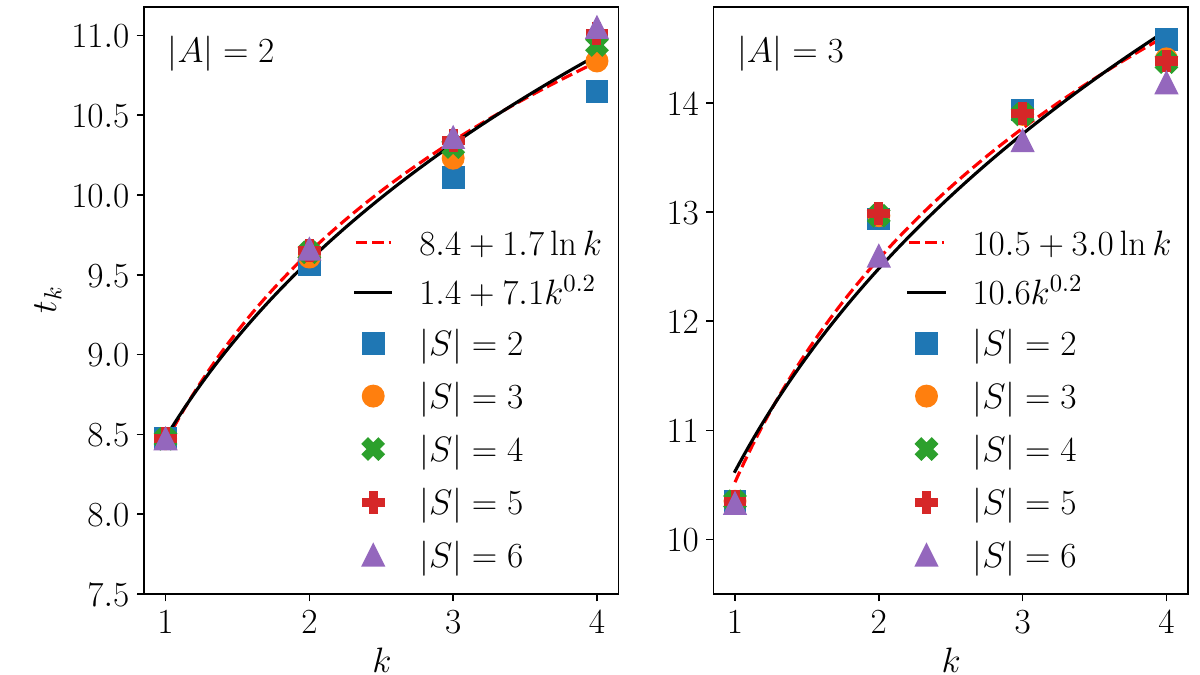}
    \caption{\new{Estimated behaviour of the deep-thermalisation timescale $t_k$ with $k$ for the kicked Ising chain. The data show no dependence of $t_k$ on $|S|$ and a slow increase with $k$, consistent with both logarithmic and sub-linear power-law growth. The timescale $t_k$ was estimated by setting a threshold $\epsilon = 3\times10^{-2}$ and identifying the time at which $\Delta_k$ drops below $\epsilon$.}}
    \label{fig:KIC-tk}
\end{figure}

The behaviour of the resulting timescale $t_k$ is shown in Fig.~\ref{fig:KIC-tk}.
As expected, we find no systematic dependence of $t_k$ on $|S|$.
By contrast, $t_k$ exhibits a slow growth with $k$.
Our results are consistent with both a logarithmic scaling $t_k \sim \ln k$ and a sublinear power-law scaling $t_k \sim k^{0.2}$; however, within the limits of our numerics, we cannot unambiguously distinguish between the two.
We note that a logarithmic dependence $t_k \sim \ln k$ is consistent with the scaling of design times observed in other toy models of deep thermalisation~\cite{ippoliti2022solvablemodelofdeep,yu2025mixedstatedeepthermalization}.

{\new{
\subsection{Finite-size scaling}
\label{sec:finite-size-scaling}

In this section, we provide additional details of the finite-size scaling analysis for $\Delta_k(t)$ for the kicked Ising chain and also the extraction of the deep-thermalisation timescale $t_k$ shown in Fig.~\ref{fig:KIC-tk}.
To start, we first show the raw data for $\Delta_k(t)$ as a function of $t$ for different system sizes $L$ and for different $|S|$, which in turn controls the mixedness of the initial state.

\begin{figure}[h]
\includegraphics[width=.75\linewidth]{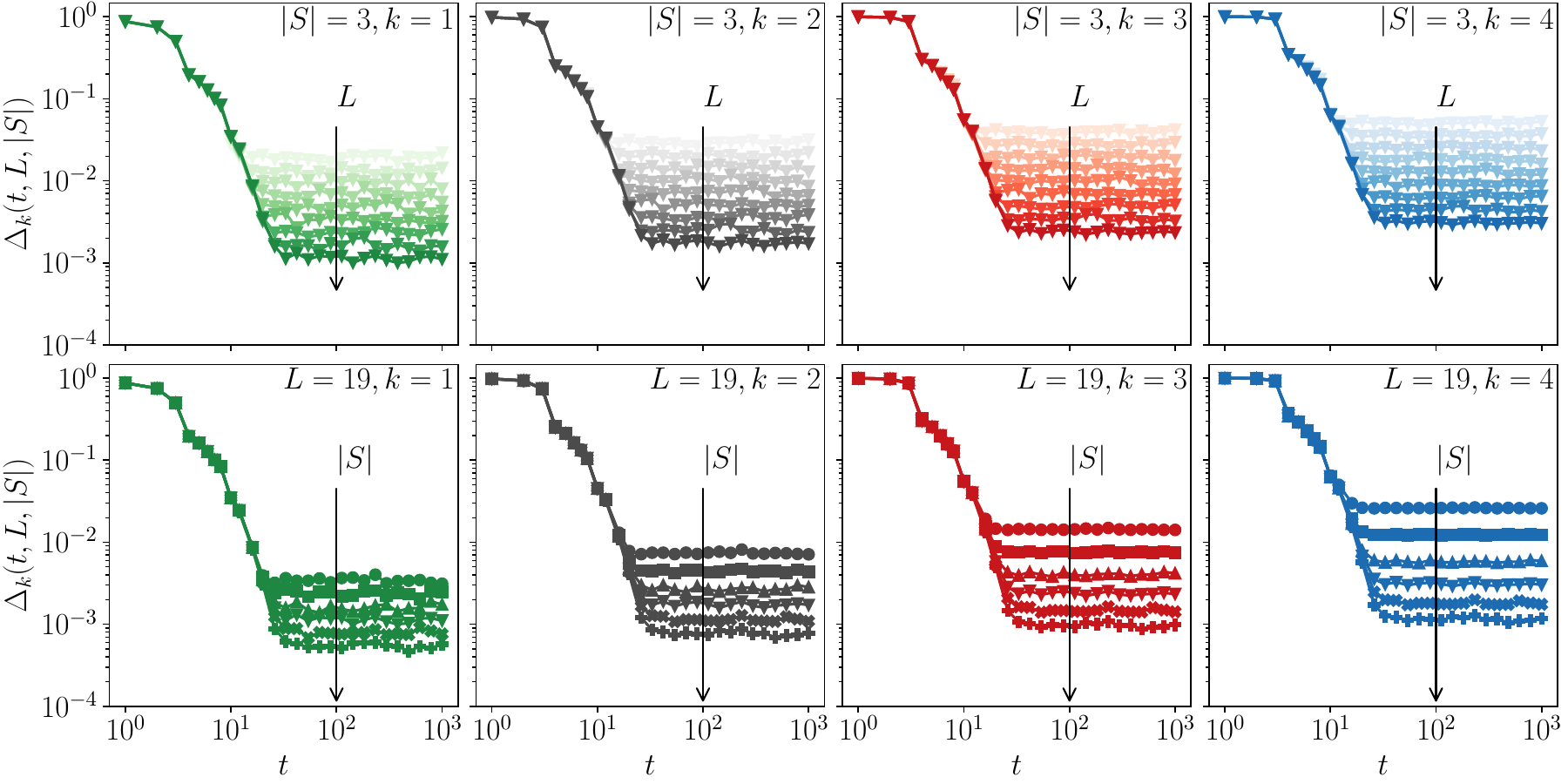}
\caption{\new{Raw data for $\Delta_k(t,L,|S|)$ in the KIC. Top panels: fixed $|S|=3$ and varying total sizes $L=11,12,\dots,19$ (increasing intensity corresponds to larger $L$, as indicated by the arrows). Bottom panels: fixed $L=19$ and varying $|S|=1,2,\dots,6$ (different symbols; trend with $|S|$ shown by the arrow).}}
\label{fig:kic-raw}
\end{figure}

The data is shown in Fig.~\ref{fig:kic-raw}. The top panels shows representative data for $\Delta_k(t)$ as a function of $t$ for fixed $|S|$ and varying total size $L$, while the bottom panels display $\Delta_k(t)$ for fixed $L$ and varying $|S|$. One immediately observes that the finite-size saturation value (which decays with $L$) depends on both $L$ and $|S|$, whereas the temporal profile prior to saturation is fully converged with them.
This motivates the scaling form
\eq{
    \Delta_k(t,L,|S|) = f_k(L,|S|)\, g_k\!\left(\frac{t}{t_{\mathrm{sat},k}(L,|S|)}\right),
    \label{eq:KIC-scaling}
}
with $g_k(x)\to 1$ for $x\gg 1$, implying
\eq{
    \Delta_k(t,L,|S|) =
    \begin{cases}
        C(t), & t\ll t_{\mathrm{sat},k}(L,|S|),\\[4pt]
        f_k(L,|S|), & t\gg t_{\mathrm{sat},k}(L,|S|),
    \end{cases}
    \label{eq:Delta-scaling-form}
}
where $t_{\mathrm{sat},k}$ denotes the finite-size saturation time.
Evidence for the validity of the scaling form in Eq.~\ref{eq:KIC-scaling} is shown by the scale-collapsed data in Fig.~\ref{fig:KIC-scaling}.
It is worth noting that $t_{\mathrm{sat},k}$ is not the deep-thermalisation timescale, which is defined in the thermodynamic limit.

\begin{figure}[t]
\includegraphics[width=0.75\linewidth]{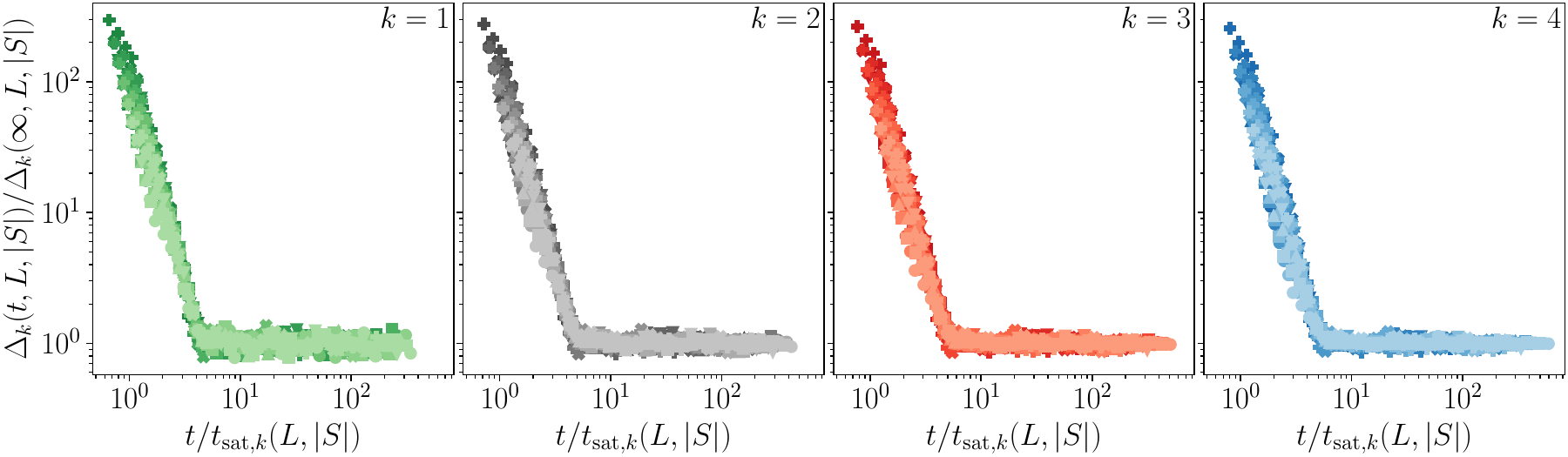}
\caption{\new{Scale collapse of the data in Fig.~\ref{fig:kic-raw} for different $L$ and $|S|$, shown separately for each $k$. The excellent quality of the collapse supports the scaling form in Eq.~\ref{eq:KIC-scaling}.}}
\label{fig:KIC-scaling}
\end{figure}

Two key conclusions follow from this analysis:
\begin{itemize}
    \item The saturation value decays exponentially with both $L$ and $|S|$,
    \eq{
        f_k(L,|S|)\equiv \Delta_k(t\!\to\!\infty,L,|S|)\sim
        e^{-\alpha_k L - \beta_k |S|}.
        \label{eq:KIC-inf-scaling}
    }
    This is evident from the data in the top panels of Fig.~\ref{fig:KIC-sat}, where $\Delta_k(\infty,L,|S|)$ decreases linearly with $|S|$ on a semilogarithmic scale, indicating exponential decay with $|S|$; the curves for linearly spaced $L$ values are similarly equispaced on logarithmic scales, indicating exponential decay with $L$ as well.
    \item The saturation time $t_{\mathrm{sat},k}$ diverges with $L$, implying that in the thermodynamic limit $\Delta_k(t,L,|S|)$ becomes independent of $|S|$. Hence, the deep-thermalisation timescale is an intrinsic property of the system and does not depend on the mixed-state partition. This is likewise evident from the bottom panels in Fig.~\ref{fig:KIC-sat}.
\end{itemize}

\begin{figure}[t]
\includegraphics[width=0.75\linewidth]{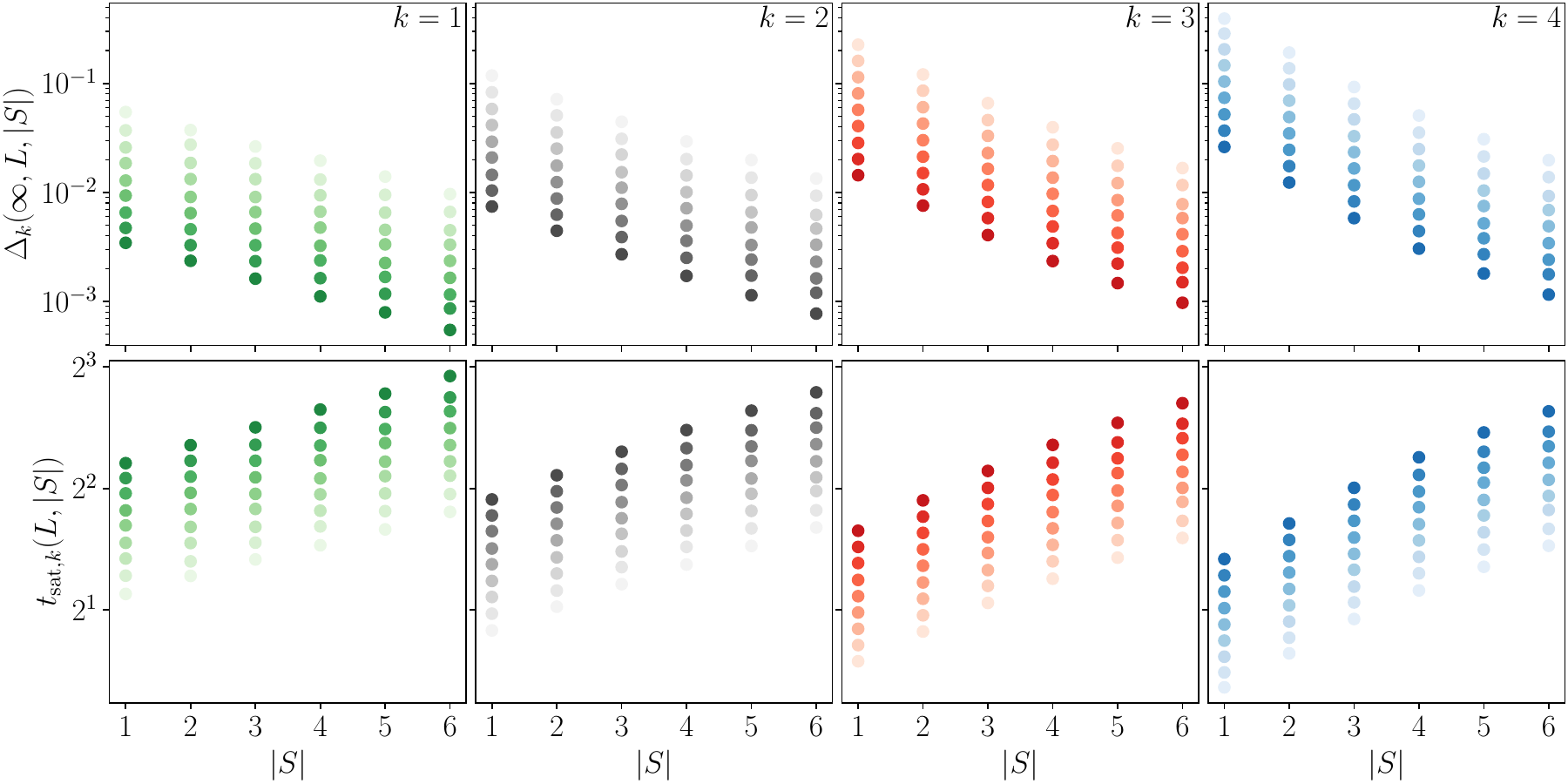}
\caption{\new{Dependence of $\Delta_k(\infty,L,|S|)$ (top) and $t_{\mathrm{sat},k}$ (bottom), which underlie the scaling collapse in Fig.~\ref{fig:KIC-scaling}, on $|S|$ and $L$. The data suggest that $\Delta_k(\infty,L,|S|)$ decays, while $t_{\mathrm{sat},k}$ grows, with both $|S|$ and $L$.}}
\label{fig:KIC-sat}
\end{figure}

To estimate $t_k$, we follow standard practice by defining it as the time when $\Delta_k$ falls below a threshold $\epsilon$. The resulting $t_k$ values, shown in Fig.~\ref{fig:KIC-tk}, exhibit a weak monotonic growth consistent with both logarithmic and sub-linear power-law scaling with $k$. However, with the current limitations on our numerical calculations, it is not possible for us to unambiguously distinguish between the two cases.

}}

\section{Calculations for the self-dual kicked Ising chain}

In this section, we detail the steps for the exact computation of the moments of the projected ensemble for the SDKI chain. 
To recall notation, the projected ensemble ${\cal E}_{\rm PE}[\rho_t]=\{p(z_B),\rho_{A,t}(z_B)\}$ is induced on subsystem $A$ by applying local projective measurements on $B$ to the state $\rho_t=U_F^t\rho_0 (U_F^\dagger)^t$.
The initial state $\rho_0$ (see Eq.~\ref{eq:rho0-kim}) is defined on a bipartition $S \cup E$ is $\rho_0=\rho_S\otimes\left(\ket{+}\bra{+}\right)^{\otimes |E|}$, where $\ket{+}$ is
the $X$-polarized state and $\rho_S$ is an arbitrary mixed state on $S$. 

It will also be useful to recall the diagrammatic notation, in Fig.~\ref{fig:kim}(a). 
At the self dual point, $J=\pi/4=h$,
$\cbox{\begin{tikzpicture}[thick,scale=0.65]
    \draw(-0.5,0) -- (0.5,0);
    \draw[fill=Orange] (-0.1,-0.1) rectangle (0.1,0.1);
\end{tikzpicture} = \begin{tikzpicture}[thick,scale=0.65]
    \draw(-0.5,0) -- (0.5,0);
    \node[draw, fill=RoyalBlue,diamond, minimum size=5,inner sep=0pt] at (0,0) {};
\end{tikzpicture} = \begin{tikzpicture}[thick,scale=0.65]
    \draw(-0.5,0) -- (0.5,0);
    \node[draw, fill=white,diamond, minimum size=5,inner sep=0pt] at (0,0) {};
\end{tikzpicture}}$ upto an irrelevant global phase with 
\eq{
\cbox{\begin{tikzpicture}[thick,scale=0.65]
    \draw(-0.5,0) -- (0.5,0);
    \node[draw, fill=white,diamond, minimum size=5,inner sep=0pt] at (0,0) {};
\end{tikzpicture}} = \frac{1}{\sqrt{2}}\begin{pmatrix}1 &1 \\1 &-1\end{pmatrix};~~~\cbox{\begin{tikzpicture}[thick,scale=0.65]\draw[fill=black] (0,0) circle (0.1);\draw(0,0) -- (0.7,0);\draw(0,0) -- (-0.5,-0.5);\draw(0,0) -- (-0.5,0.5);\node at (0.95,0){$z_1$};\node at (-0.7,0.5){$z_2$};\node at (-0.7,-0.5){$z_3$};\node at (0,-0.5){$g$};\end{tikzpicture}} = \delta{z_1z_2z_3}e^{-ig(1-2z_1)}\,.\label{eq:sdki-motif}
}

Consider the set of unnormalised mixed states in the projected ensemble, given by $\{\tilde{\rho}_{A,t}(z_B):=p(z_B)\rho_{A,t}(z_B)\}$. 
One can consider a set of states, $\{\ket{\tilde{\psi}_{XA,t}(z_B)}\}$, which is a purification of the set $\{\tilde{\rho}_{A,t}(z_B)\}$ satisfying $\tilde{\rho}_{A,t}(z_B)=\text{Tr}_X\left[\ket{\tilde{\psi}_{XA,t}(z_B)}\bra{\tilde{\psi}_{XA,t}(z_B)}\right]$, where $X$ is an auxiliary subsystem that is traced over.
The diagrammatic notation in Eq.~\ref{eq:sdki-motif} allows us to represent the state $\ket{\tilde{\psi}_{XA,t}(z_B)}$ as a tensor network, given by
\eq{
\ket{\tilde{\psi}_{XA,t}(z_B)} &= \sqrt{2}^{t(|A|+|B|-1)} \times \cbox{
\begin{tikzpicture}[thick,scale=0.8, transform shape]
\draw[rounded corners=3pt, color=RoyalPurple, fill=RoyalPurple!5] (-0.2,-0.2) rectangle (8.2,0.75);
\foreach \x in {0,...,8}{
    \draw(\x,0) -- (\x,1);
    \draw(\x,2) -- (\x,4);
    \draw(\x,-0.45) -- (\x,0);
    \draw[dashed](\x,1) -- (\x,2);
  \foreach \y in {0,...,3}{
    \draw[fill=black] (1*\x,1*\y) circle (0.08);
    \draw(0,\y) -- (1,\y);
    \draw[dashed](1,\y) -- (2,\y);
    \draw(2,\y) -- (6,\y);
    \draw[dashed](6,\y) -- (7,\y);
    \draw(7,\y) -- (8,\y);
    }
}
\foreach \x in {0,...,8}{
\foreach \y in {0,2,3}{
\node[draw, fill=white,diamond, minimum size=5,inner sep=0pt] at (\x,\y+0.5) {};
}}
\foreach \y in {0,...,3}{
\node[draw, fill=white,diamond, minimum size=5,inner sep=0pt] at (0.5,\y) {};
\node[draw, fill=white,diamond, minimum size=5,inner sep=0pt] at (7.5,\y) {};
\foreach \x in {2,...,5}{
\node[draw, fill=white,diamond, minimum size=5,inner sep=0pt] at (\x+0.5,\y) {};
}
}
\draw[rounded corners=3pt, fill=black!20] (-0.2,-1.05) rectangle (3.2,-0.45);
\node at (1.5, -0.75) {$\sqrt{\rho_S}$};
\foreach \x in {0,...,3}{
\draw(\x,-1.05) -- (\x,-1.15-0.08*\x);
\draw (\x,-1.15-0.08*\x) arc (0:-90:0.1);
\draw (\x-0.1,-1.15-0.08*\x-0.1) -- (-0.5-0.4*\x,-1.15-0.08*\x-0.1);
\draw(-0.5-0.4*\x,-1.15-0.08*\x-0.1) arc (270:180:0.1);
\draw(-0.5-0.4*\x-0.1,-1.15-0.08*\x-0.1+0.1) -- (-0.5-0.4*\x-0.1,4);
}
\foreach \x in {4,...,8}{
\filldraw[fill=blue!20] (\x-0.2,-0.45) -- (\x+0.2,-0.45) -- (\x,-0.65) -- cycle;
\filldraw[fill=red!20] (\x-.2,4) -- (\x+0.2,4) -- (\x,4.2) -- cycle;
}
\filldraw[fill=red!20] (3-.2,4) -- (3+0.2,4) -- (3,4.2) -- cycle;
\draw[dashed,Bittersweet](2.2,-0.45)--(2.2,4.3);
\draw [thick, decorate, decoration={brace,amplitude=5pt,mirror,raise=4pt},yshift=0pt]
(8.1,0) -- (8.1,3.5) node [black,midway,xshift=15pt] {$t$};
\draw [thick, decorate, decoration={brace,amplitude=5pt, raise=3pt},yshift=0pt]
(0,4.1) -- (2,4.1) node [black,midway,yshift=15pt] {$A$};
\draw [thick, decorate, decoration={brace,amplitude=5pt, raise=3pt},yshift=0pt]
(3,4.1) -- (8,4.1) node [black,midway,yshift=15pt] {$B$};
\draw [thick, decorate, decoration={brace,amplitude=5pt, raise=3pt},yshift=0pt]
(-1.8,4.1) -- (-0.6,4.1) node [black,midway,yshift=15pt] {$X$};
\end{tikzpicture}
}\,,}
where the blue downward pointing triangles on denote the $\ket{+}$ state and the red upward pointing triangles denote the Pauli-$Z$ (computational basis) measurement. 
In terms of these unormmalise states, the $k^{th}$ moments of ${\cal E}_{\rm PE}[\rho_t]$ is given by,
\begin{equation}	
\rho^{(k)}_{{\cal E}_{\rm PE}}=\sum_{z_B}\tfrac{\left(\text{Tr}_X\left[\ket{\tilde{\psi}_{XA,t}(z_B)}\bra{\tilde{\psi}_{XA,t}(z_B)}\right]\right)^{\otimes k}}{\braket{\tilde{\psi}_{XA,t}(z_B)|\tilde{\psi}_{XA,t}(z_B)}^{k-1}}\,.
	\label{eq:moments-unnorm}
\end{equation}

A key feature of the SDKI chain is that the measurements in $B$ can be traded for a sequence of unitary operators $U(z_{B_i})$, which explicitly depend on the measurement outcome, acting along the spatial direction with the boundary condition being $\ket{+}^{\otimes t}$~\cite{ho2022exact}.
Diagrammatically, 
\eq{\ket{\tilde{\psi}_{XA,t}(z_B)} 
&=
\frac{1}{\sqrt{2}^{|B|-|S|}} \times \cbox{
\begin{tikzpicture}[thick,scale=0.8, transform shape]
\foreach \x in {0,...,8}{
    \draw(\x,0) -- (\x,1);
    \draw(\x,2) -- (\x,3);
    \draw[dashed](\x,1) -- (\x,2);
  \foreach \y in {0,...,3}{
    \draw[fill=black] (1*\x,1*\y) circle (0.08);
    \draw(0,\y) -- (1,\y);
    \draw[dashed](1,\y) -- (2,\y);
    \draw(2,\y) -- (6,\y);
    \draw[dashed](6,\y) -- (7,\y);
    \draw(7,\y) -- (8,\y);
    }
}
\foreach \x in {0,...,2}{
\draw(\x,3) -- (\x,5);
\node[draw, fill=white,diamond, minimum size=5,inner sep=0pt] at (\x,3+0.5) {};
}
\foreach \x in {0,...,8}{
\foreach \y in {0,2}{
\node[draw, fill=white,diamond, minimum size=5,inner sep=0pt] at (\x,\y+0.5) {};
}}
\foreach \y in {0,...,3}{
\node[draw, fill=white,diamond, minimum size=5,inner sep=0pt] at (0.5,\y) {};
\node[draw, fill=white,diamond, minimum size=5,inner sep=0pt] at (7.5,\y) {};
\foreach \x in {2,...,5}{
\node[draw, fill=white,diamond, minimum size=5,inner sep=0pt] at (\x+0.5,\y) {};
\draw(8,\y)--(8.35,\y);
\filldraw[fill=blue!20] (8.35,\y-0.2) -- (8.35,\y+0.2) -- (8.55,\y) -- cycle;
}
}
\foreach \x in {0,...,3}{
\draw(\x,0) -- (\x,-1.35);
}
\draw[rounded corners=3pt, fill=black!20] (-0.2,-1.05) rectangle (3.2,-0.45);
\node at (1.5, -0.75) {$\sqrt{\rho_S}$};
\foreach \x in {3,4,5,7}{
\draw[rounded corners=3pt, color=YellowOrange] (\x+0.25,-0.7) rectangle (\x+1+0.2,4.5);
\node[rotate=90] at (\x+0.6, 3.8) {$\times \sqrt{2}^{t-1}$};
}
\node[rotate=90] at (2+0.6, 3.8) {$\times \sqrt{2}^{t-1}$};
\node at (4+0.725, -0.5) {\small {\color{YellowOrange}${U}\!(z_{B_i})$}};
\draw[rounded corners=3pt, color=Red] (-2,-0.2) rectangle (3.15,4.5);
\node at (-1., 2) {$\sqrt{2}^{|A|(t-1)}\times$};
\node at (-1.25, 0.15) { {\color{Red}${W(z_{B \cap S})}$}};
\draw[dashed,Bittersweet](2.2,-0.45)--(2.2,4.3);
\draw [thick, decorate, decoration={brace,amplitude=5pt,mirror,raise=4pt},yshift=0pt]
(8.5,-0.2) -- (8.5,3.3) node [black,midway,xshift=15pt] {$t$};
\draw [thick, decorate, decoration={brace,amplitude=5pt, raise=3pt},yshift=0pt]
(0,5.1) -- (2,5.1) node [black,midway,yshift=15pt] {$A$};
\end{tikzpicture}
}.
}
Defining the product of the unitaries enclosed in the yellow boxes in the above equation as, 
\eq{\mathcal{U}(z_{B\cap E}):= U(z_{1})U(z_{2})...U(z_{|B\cap E|})\,,}
the unnormalised state can be written as 
\eq{\ket{\tilde{\psi}_{XA,t}(z_B)}=\frac{1}{\sqrt{2}^{|B|-|S|}}\left(\sqrt{\rho_S}^T\otimes I_A\right)W(z_{B\cap S})~\mathcal{U}(z_{B\cap E})\ket{+}^{\otimes t}\,,} 
where $W(z_{B\cap S})$ is a linear map from $\mathbb{C}^{2^t}\rightarrow\mathbb{C}^{2^{|A|+|S|}}$. Plugging this into Eq.~\ref{eq:moments-unnorm}, we obtain
\begin{equation}
	\rho^{(k)}_{{\cal E}_{\rm PE}}=\sum_{z_B}\frac{1}{2^{|B|}}\tfrac{\left(\text{Tr}_X\left[\left(\sqrt{\rho_S}^T\otimes I_A\right)W(z_{B\cap S})\mathcal{U}(z_{B\cap E})\left(\ket{+}\bra{+}^{\otimes t}\right) \mathcal{U}(z_{B\cap E})^\dagger W(z_{B\cap S})^\dagger\left(\sqrt{\rho_S}^T\otimes I_A\right)\right]\right)^{\otimes k}}{\left(\bra{+}^{\otimes t}\mathcal{U}(z_{B\cap E})^\dagger W(z_{B\cap S})^\dagger \left(\rho_S^T\otimes I_A\right) W(z_{B\cap S}) \mathcal{U}(z_{B\cap E})\ket{+}^{\otimes t}\right)^{k-1}}\times 2^{|S|}\,.
	\label{eq:DU-moments}
\end{equation}

For $t\geq |A|+|S|$, the map $W(z_{B\cap S})$ is expressible as 
\eq{W(z_{B\cap S})=\sqrt{2}^{(t-|A|-|S|)}\bra{+}^{\otimes^{(t-|A|-|S|)}}V(z_{B\cap S})\,,} where $V(z_{B\cap S})$ is a unitary on $\mathbb{C}^{2^t}$ whose particular form is unimportant. This is made explicit by the diagrammatic notation 
\eq{
\cbox{\begin{tikzpicture}[thick, scale=0.8]
\foreach \y in {0,...,8}{
\draw[dashed](1,\y)--(2,\y);
\foreach \x in {0,...,3}{
\draw[fill=black] (\x,\y) circle (0.08);
\draw[dashed](\x,6)--(\x,7);
}}  
\foreach \y in {0,1,2,3,4,5,7,8}{
\draw(1,\y)--(1,\y+1);
\node[draw, fill=white,diamond, minimum size=5,inner sep=0pt] at (1,\y+0.5) {};
\foreach \x in {0,2}{
\draw(\x,\y)--(\x+1,\y);\draw(\x,\y)--(\x,\y+1);
\node[draw, fill=white,diamond, minimum size=5,inner sep=0pt] at (\x+0.5,\y) {};
\node[draw, fill=white,diamond, minimum size=5,inner sep=0pt] at (\x,\y+0.5) {};
}}
\foreach \y in {0,1,2,3,4,5,7}{
\draw(3,\y)--(3,\y+1);
\node[draw, fill=white,diamond, minimum size=5,inner sep=0pt] at (3,\y+0.5) {};}
\draw(0,6)--(1,6);\draw(2,6)--(3,6);
\foreach \x in {0,...,3}{
\draw(\x,0)--(\x,-0.5);
}
\foreach \y in {0,...,8}{
\draw(3,\y)--(3.5,\y);
}
\draw[dashed,color=Bittersweet](2.6,8.4)--(-0.4,5.4);
\draw[dashed,color=CornflowerBlue](3.25+0.1,-0.25+0.1)--(-0.5+0.1,3.5+0.1);
\node at (-1.5, 4.5) {$\sqrt{2}^{|S|(t-1)}\times$};
\draw[rounded corners=3pt, color=Red] (-2.6,-0.2) rectangle (3.15,8.7);
\node at (-1.5, 0.1) {{\color{Red}$W(z_{B\cap S})$}};
\draw [thick, decorate, decoration={brace,amplitude=5pt,mirror,raise=4pt},yshift=0pt]
(3.5,-0.2) -- (3.5,8.3) node [black,midway,xshift=15pt] {$t$};
\draw [thick, decorate, decoration={brace,amplitude=5pt, raise=3pt},yshift=0pt](0,9) -- (2,9) node [black,midway,yshift=15pt] {$A$};
\draw [thick, decorate, decoration={brace,mirror,amplitude=5pt, raise=3pt},yshift=0pt](-0.1,-0.35) -- (3.1,-0.35) node [black,midway,yshift=-15pt] {$S$};
\node[draw, fill=white,diamond, minimum size=5,inner sep=0pt] at (0.5,6) {};
\node[draw, fill=white,diamond, minimum size=5,inner sep=0pt] at (2.5,6) {};
\end{tikzpicture}
}
=\sqrt{2}^{t-|A|-|S|}\times
\cbox{
\begin{tikzpicture}[thick,scale=0.8]
    \foreach \y in {8,7,6,5,4,3,2,1}{
    \ifthenelse{\y=7}{\draw[dashed](3,\y)--(3,\y-1);
    \draw[fill=black] (3,\y) circle (0.08);}{
    \draw(3,\y)--(3,\y-1);
    \draw[fill=black] (3,\y) circle (0.08);
    \node[draw, fill=white,diamond, minimum size=5,inner sep=0pt] at (3,\y-0.5) {};}
    }
    \foreach \y in {8,7,6,5,4,3,2}{
    \ifthenelse{\y=7}{\draw[dashed](2,\y)--(2,\y-1);}{\draw(2,\y)--(2,\y-1);\node[draw, fill=white,diamond, minimum size=5,inner sep=0pt] at (2,\y-0.5) {};}
    \ifthenelse{\y < 8}{\draw[fill=black] (2,\y) circle (0.08);}{}
    }
    \foreach \y in {7,6,5,4,3}{
    \ifthenelse{\y=7}{\draw[dashed](1,\y)--(1,\y-1);}{\draw(1,\y)--(1,\y-1);\node[draw, fill=white,diamond, minimum size=5,inner sep=0pt] at (1,\y-0.5) {};}
    \ifthenelse{\y < 7 }{\draw[fill=black] (1,\y) circle (0.08);}{}
    }
    \foreach \y in {6,5,4}{
     \ifthenelse{\y < 6 }{\draw[fill=black] (0,\y) circle (0.08);}{}
    \draw(0,\y)--(0,\y-1);
    \node[draw, fill=white,diamond, minimum size=5,inner sep=0pt] at (0,\y-0.5) {};
    }
    \foreach \y in {0,1,2,3}{
    \draw(3-\y,\y)--(-4,\y);}
    \foreach \i in {7,8}{
    \draw(\i-6,\i)--(-4,\i);
    }
    \draw(0,6)--(-1,6);\draw[dashed](-1,6)--(-2,6);\draw(-4,6)--(-2,6);
    \foreach \i in {0,...,3}{
    \draw[fill=black] (0-\i,3) circle (0.08);
    \ifthenelse{\i<3}{\node[draw, fill=white,diamond, minimum size=5,inner sep=0pt] at (0-\i-0.5,3) {};}{}
    }
    \foreach \i in {1,...,3}{
    \draw[fill=black] (-\i,2) circle (0.08);
    \draw[fill=black] (-\i,6) circle (0.08);
    \ifthenelse{\i<3}{\node[draw, fill=white,diamond, minimum size=5,inner sep=0pt] at (0-\i-0.5,2) {};}{}
    \draw(-\i,3)--(-\i,2);
    }
    \draw(-3,0)--(-3,1);
    \foreach \x in {-2,-3} \draw[dashed](\x,1)--(\x,2);
    \foreach \i in {2,...,3}{
    \draw[fill=black] (0-\i,1) circle (0.08);
    \draw[fill=black] (-\i,7) circle (0.08);
    \node[draw, fill=white,diamond, minimum size=5,inner sep=0pt] at (0-\i-0.5,7) {};
    \node[draw, fill=white,diamond, minimum size=5,inner sep=0pt] at (0-\i-0.5,6) {};
    \ifthenelse{\i<3}{\node[draw, fill=white,diamond, minimum size=5,inner sep=0pt] at (0-\i-0.5,1) {};}{}
    }
    \foreach \i in {3}{
    \draw[fill=black] (0-\i,0) circle (0.08);
    \draw[fill=black] (-\i,8) circle (0.08);
    \node[draw, fill=white,diamond, minimum size=5,inner sep=0pt] at (0-\i-0.5,8) {};
    \ifthenelse{\i<3}{\node[draw, fill=white,diamond, minimum size=5,inner sep=0pt] at (0-\i-0.5,2) {};}{}
    }
    \foreach \i in {1,3}{
    \node[draw, fill=white,diamond, minimum size=5,inner sep=0pt] at (-3,4-\i-0.5) {};}
    \foreach \i in {1,2}{
    \ifthenelse{\i=1}{\draw(-4+\i,6)--(-4+\i,9-\i-1);\draw[dashed](-4+\i,6+1)--(-4+\i,9-\i);\node[draw, fill=white,diamond, minimum size=5,inner sep=0pt] at (-2,4-\i-0.5) {};}{\draw(-4+\i,6)--(-4+\i,9-\i);}
    }
    \node[draw, fill=white,diamond, minimum size=5,inner sep=0pt] at (-3,6.5) {};
    \node[draw, fill=white,diamond, minimum size=5,inner sep=0pt] at (-1,2.5) {};
    \node[draw, fill=white,diamond, minimum size=5,inner sep=0pt] at (-2,6.5) {};
    \draw[dashed,color=Bittersweet](2.6,8.4)--(-0.4,5.4);
\draw[dashed,color=CornflowerBlue](3.25+0.1,-0.25+0.1)--(-0.5+0.1,3.5+0.1);
\foreach \y in {0,...,8}{
\ifthenelse{\y=0}{\draw(3,\y)--(3.5,\y);}{
\draw(2,\y)--(3.5,\y);
\node[draw, fill=white,diamond, minimum size=5,inner sep=0pt] at (2.5,\y) {};}
}
\foreach \y in {3,...,6}{
\draw(0,\y)--(1,\y);
\node[draw, fill=white,diamond, minimum size=5,inner sep=0pt] at (0.5,\y) {};}
\foreach \y in {2,...,7}{
\draw[dashed](1,\y)--(2,\y);}
\foreach \y in {4,5}{ \draw[](0,\y)--(-4.75,\y);
\filldraw[fill=blue!20] (-4.75,\y-0.2) -- (-4.75,\y+0.2) -- (-4.95,\y) -- cycle;}
\node at (4.5, 4.5) {$\times \sqrt{2}^{|S|(t-1)}$};
\foreach \i in {0,1,2,3,6,7,8}\draw(-4,\i)--(-4.8,\i);
\draw [thick, decorate, decoration={brace,amplitude=5pt,mirror,raise=4pt},yshift=0pt](5.9,-0.2) -- (5.9,8.4) node [black,midway,xshift=15pt] {$t$};
\draw [thick, decorate, decoration={brace,amplitude=5pt,raise=4pt},yshift=0pt](-4.7,-0.2) -- (-4.7,3.2) node [black,midway,xshift=-15pt] {$S$};
\draw [thick, decorate, decoration={brace,amplitude=5pt,raise=4pt},yshift=0pt](-4.7,5.8) -- (-4.7,8.2) node [black,midway,xshift=-15pt] {$A$};
\draw[rounded corners=3pt, color=Green] (-4.5,-0.7) rectangle (5.9,8.4);
\node at (-3.4, -0.4) {{\color{Green}$V(z_{B\cap S})$}};
\end{tikzpicture}}\,.}
The diagram on the right hand side of the above equation follows trivially from the left hand side by rotating the wordlines and the gates about the corresponding dashed (red and blue) lines. 

The second key feature of the SDKI is that the products of unitary matrices $U_{z_{B_i}}$ sample uniformly the space of unitaries in $\mathbb{C}^{2^t}$~\cite{ho2022exact}, which enable the analytic computation of the moments.
In particular, in the limit of $|B\cap E|\to\infty$, Theorem 2 in Ref.~\cite{ho2022exact} states that $\forall ~g/\in Z\pi/8$ 
\eq{\lim_{|B\cap E|\rightarrow\infty} \sum_{z_{B\cap E}}
\frac{1}{2^{|B\cap E|}}\mathcal{U}(z_{B\cap E})^{\otimes k} \otimes \mathcal{U}(z_{B\cap E})^
{*\otimes k} =\int dU_{\text{Haar}(2^t)}U^{\otimes k}\otimes U^{*\otimes k}\,.
}
This enables us to replace the sum over states $\mathcal{U}(z_B)\ket{+}^{\otimes t}$ in Eq.~\ref{eq:DU-moments} with the integral over the Haar random states $\Psi\sim\text{Haar}(2^t)$. The unitary $V(z_{B\cap S})$ can be absorbed in the Haar integral via invariance of the Haar measure. Defining $\ket{\Psi_+}:=\bra{+}^{\otimes (t-|A|-|S|)}\ket{\Psi}$ and $\ket{\Psi'}:=\frac{\ket{\Psi_+}}{|\ket{\Psi_+}|}$ and $\rho_{XA}=\rho_S^T\otimes I_A/2^{|A|}$, the moments in the $|B\cap E|\to\infty$ limit can be conveniently expressed as 
\begin{equation} 
	\begin{split}
	\lim_{|B\cap E|\rightarrow\infty}\rho^{(k)}_{{\cal E}_{\rm PE}}=&\int d\Psi_{\text{Haar}(2^t)}\frac{\left(\text{Tr}_X\left[\left(\sqrt{\rho_{XA}}\right)\ket{\Psi_+}\bra{\Psi_+}\left(\sqrt{\rho_{XA}}\right)\right]\right)^{\otimes k}}{\braket{\Psi_+|\left(\rho_{XA}\right)|\Psi_+}^{k-1}}\times~2^{t}\\
		=&\int d\Psi_{\text{Haar}(2^t)}\frac{\left(\text{Tr}_{X}\left[\sqrt{\rho_{XA}}\ket{\Psi'}\bra{\Psi'}\sqrt{\rho_{XA}}\right]\right)^{\otimes k}}{\braket{\Psi'|\rho_{XA}|\Psi'}^k}\times2^t\braket{\Psi'|\rho_{XA}|\Psi'}\times \braket{\Psi_+|\Psi_+}\,.
		\label{eq:DU-moment-rough}
	\end{split}
\end{equation}
Lemma 4 of~\cite{cotler2023emergent} states that $\ket{\Psi'}$ and $\braket{\Psi_+|\Psi_+}$ are independent random variables, yielding
\begin{equation}
	\begin{split}
		\lim_{|B\cap E|\rightarrow\infty}\rho^{(k)}_{{\cal E}_{\rm PE}}=&\int d\Psi'_{\text{Haar}(2^t)}\frac{\left(\text{Tr}_{X}\left[\sqrt{\rho_{XA}}\ket{\Psi'}\bra{\Psi'}\sqrt{\rho_{XA}}\right]\right)^{\otimes k}}{\braket{\Psi'|\rho_{XA}|\Psi'}^k}\times2^{|A|+|S|}\braket{\Psi'|\rho_{XA}|\Psi'} \\
		&\times\int d\Psi_{\text{Haar}(2^t)} ~2^{t-|A|-|S|}\braket{\Psi_+|\Psi_+}
		\label{eq:DU-moment-final}
	\end{split}
\end{equation}
where the latter integral is unity, thus giving the result in the main text.

{\new{
\section{Results for the Floquet mixed-field Ising chain}
In the main text, our results, both numerical and analytical, were presented for the non-integrable kicked Ising chain (KIC) and at its self-dual point. While the non-integrable KIC is an archetypal model for chaotic/ergodic unitary dynamics, for the sake of generality (and completeness) we also present results here for the Floquet mixed-field Ising chain (FMFIC) with the measurements performed in a basis other than the computational basis. The Floquet unitary of the FMFIC is as
\begin{equation}
U_{F} = e^{-iH_+}e^{-iH_-}, \quad {\rm with} \quad H_{\pm} = \sum_{i=1}^N [h_z(1\pm\delta)Z_i + h_xX_i + J Z_iZ_{i+1}],.
\label{eq:FMFIM}
\end{equation}
In this model, both steps of the Floquet period include Ising interactions as well as {\it both} transverse and longitudinal fields. As such, it cannot be recast as a reparametrisation of the KIC. Moreover, the projected ensemble is generated by measurements of $\{Y_i : i \in B\}$, which are, by construction, not in the computational basis.
\begin{figure}[h]
\centering
\includegraphics[width=0.4\linewidth]{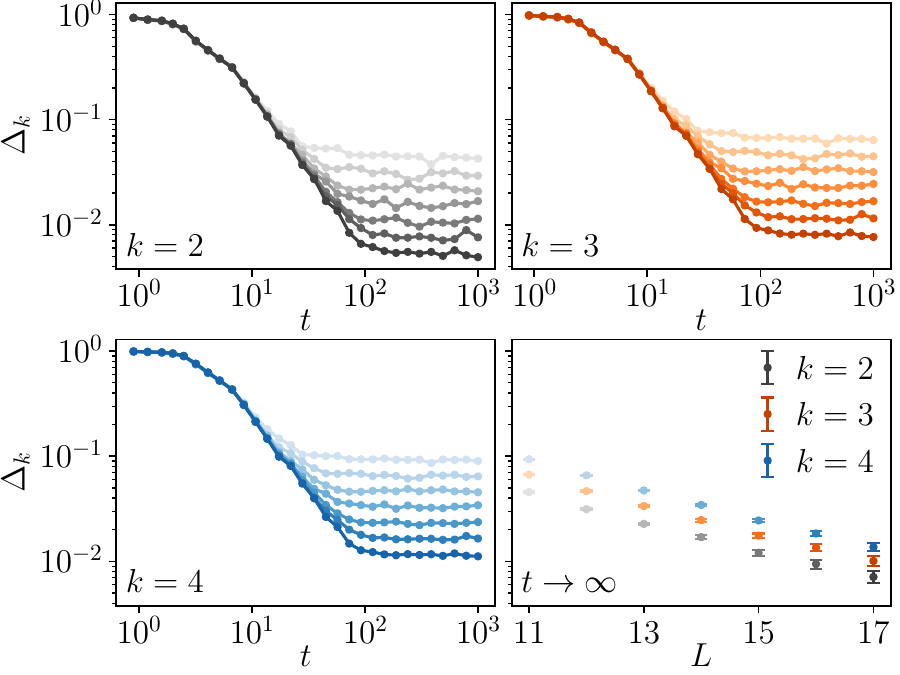}
\caption{{\new{Results for $\Delta_k(t)$ for the FMFIC with parameters $(h_z,h_x,J,\delta,T)=(0.8090, 0.9045,1,0.5,1)$. The subsystem size $|A|=3$ is fixed, while different intensities correspond to $|B|=8,10,\dots,14$, with darker colours denoting larger values. The initial state is of the form in Eq.~\ref{eq:rho0-kim} with $\rho_S$ a maximally mixed state with $|S|=3$. The results are, qualitatively, indistinguishable from those for the KIC shown in Fig.~\ref{fig:kim} of the main text.}}}
\label{fig:mfim}
\end{figure}

We again start from an initial state of the form in Eq.~\ref{eq:rho0-kim} where we take $\rho_S$ to be the maximally mixed state and $\ket{v_j}$ to be the eigenstate of $Y_j$ with eigenvalue $+1$. However, given the ergodic nature of the FMFIC, the choice of the state $\ket{v_j}$ is immaterial.
The results for $\Delta_k(t)$ are shown in Fig.~\ref{fig:mfim}. The behaviour of $\Delta_k(t)$ as a function of time $t$ and system size $L$ is qualitatively indistinguishable from that of the KIC, thus demonstrating the general robustness of our results.

\section{Models with weak ergodicity breaking: PXP model}

Throughout this work, our motivation has been to understand the physics of mixed-state deep thermalisation under chaotic/ergodic unitary dynamics. On the other hand, the question of the fate of mixed-state deep thermalisation in the case of ergodicity broken systems is certainly an important and interesting extension of this. 
A comprehensive exploration of this topic lies beyond the scope of the present work. However, as proof of principle we present some preliminary numerical reults in this section for a model with weak ergodicity breaking.

Specifically, we consider a version of the of PXP model, described by the Hamiltonian
\begin{equation}
    H = H_{\mathrm{PXP}} + H_\mu = \sum_j P_{j-1} X_j P_{j+1} + \mu \sum_j Z_j\,, 
    \qquad P_j = (1 - Z_j)/2\,,
    \label{eq:HPXP}
\end{equation}
which is an archetypal model for weak ergodicity breaking via quantum many-body scars.
For this model, Ref.~\cite{bhore2023deepthmconstrained} found that while all initial states ultimately deeply thermalise, scarring states exhibit significantly longer deep-thermalisation timescales.

To adapt this setting to our mixed-state framework, we considered initial states of the form Eq.~\ref{eq:rho0-kim} (in the main text), with $\rho_S \propto \mathbb{I}_{D_S}$ and the pure-state component $\prod_{j \in E} |v_j\rangle\langle v_j|$ chosen to be either a scarring state ($|Z_2\rangle$) or a non-scarring state ($|Z_4\rangle$). Here $\ket{Z2} = \ket{\uparrow\downarrow\uparrow\downarrow\cdots \uparrow\downarrow}$ and $\ket{Z4}=\ket{\uparrow\downarrow\downarrow\downarrow\uparrow\downarrow\downarrow\downarrow\cdots \uparrow\downarrow\downarrow\downarrow}$.

We then examined the dynamics of the deep-thermalisation measure $\Delta_k(t)$.
The results are shown in Fig.~\ref{fig:pxp} for different values of $|S|$.

\begin{figure}[h]
\centering
\includegraphics[width=0.6\linewidth]{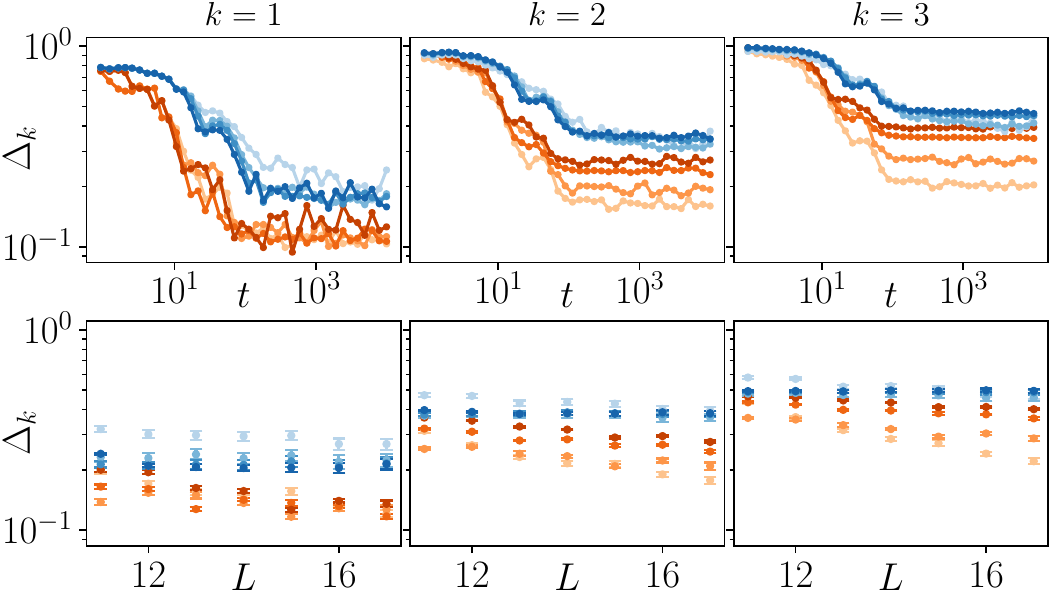}
\caption{\new{
Results for $\Delta_k(t)$ for the PXP model defined in Eq.~\ref{eq:HPXP} with $|A| = 3$ and $\mu = 0.05$. 
The initial state is of the form in Eq.~\ref{eq:rho0-kim} of the main text, with $S$ in a maximally mixed state and $E$ either $|Z_2\rangle$ (blue) or $|Z_4\rangle$ (red). 
Different intensities correspond to subsystem sizes $|S| = 0, 2, 4, 6$ (tuning the mixedness), with darker colours denoting larger $|S|$. 
Upper panels: time evolution of $\Delta_k(t)$ for total system size $L=17$. 
Lower panels: long-time average as a function of $L$.}}
\label{fig:pxp}
\end{figure}

The results show that the overall phenomenology is robust upon introducing mixed states: the scarring state takes significantly longer to deep-thermalise than the non-scarring one, confirming that the mixed-state version remains a sensitive probe of weak ergodicity breaking. 
At the same time, increasing the rank of the mixed part of the initial state systematically enhances the deep-thermalisation timescale. 
This behaviour arises because a mixed initial state explores a larger portion of Hilbert space, thereby sampling more extensively from the family of scarred eigenstates. 
These effects are most pronounced for $k \ge 2$ and become more marked with increasing $k$.

Hence, our results can be interpreted in two complementary ways. 
First, mixed-state deep thermalisation continues to clearly distinguish between scarring and non-scarring initial states, particularly at low rank of the initial mixed state. 
Second, increasing the mixedness of the initial state amplifies sensitivity to weak ergodicity breaking across the spectrum---potentially providing a practical diagnostic tool for detecting hidden scarring in other models.

}}

\section{Probabilities of (bitstring) probabilities}

As an experimental and operational probe of mixed-state deep thermalisation we present some results for the probabilities of (bitstring) probabilities (PoPs) over the projected ensemble. 
Formally, they are defined over the PE in Eq.~\ref{eq:PE-def}, for a bit string $z_A$, as 
\eq{
{\rm PoP}_{z_A}(\tilde{p}) = \sum_{z_B}p(z_B)\delta(\tilde{p} - \tilde{p}(z_A|z_B))\,,
}
where $p(z_A|z_B) = {\rm Tr} [\ket{z_A}\bra{z_A}\varrho_A(z_B)]$ is the probability of obtaining the bitstring $z_A$ in the conditional state $\varrho_A(z_B)$ and the $\tilde{p}(z_A|z_B) = p(z_A|z_B)/\braket{p(z_A)}$ denotes the relative probability with $\braket{p(z_A)} = \sum_{z_B}p(z_B) p(z_A|z_B)$.

\begin{figure}[t]
\includegraphics[width=.5\linewidth]{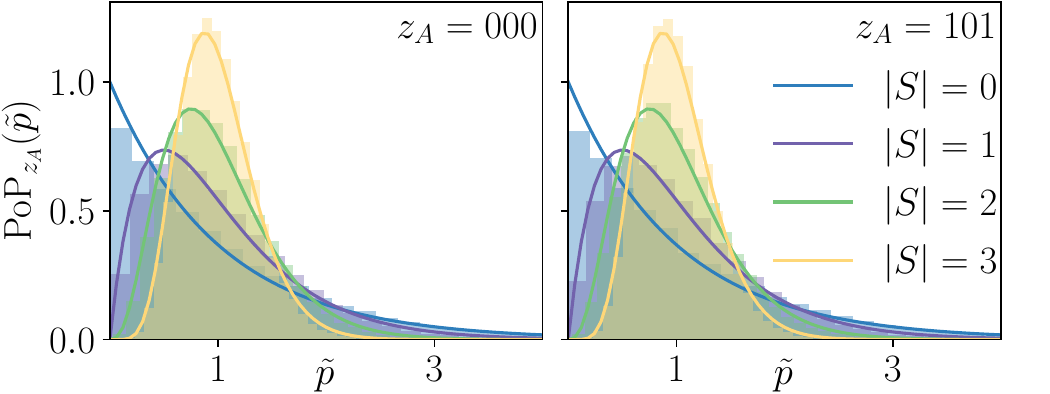}
    \caption{{\new {The PoPs over the projected ensemble for the kicked Ising chain at a late time of $t=1000$ for $L=19$ and $|A|=3$. Different colours correspond to different $|S|=0,1,2,3$ (which also tunes the entropy of the initial state). The histograms correspond to the numerical data whereas the solid lines denote the Erlang distribution (Eq.~\ref{eq:erlang}) for the corresponding $D_S$. The two panels denote two different bitstrings in $A$.}}}
    \label{fig:kic-pop}
\end{figure}

The results are shown for the kicked Ising chain at a late time of $t=10^3$, for two respresentative bitstrings $z_A=000$ and $z_A=101$,  in Fig.~\ref{fig:kic-pop}. The results are for the case when $\rho_S$ in the initial state in Eq.~\ref{eq:rho0-kim} was chosen to be maximally mixed.

Remarkably, the PoPs for any non-zero $|S|$ are extremely well described by the Erlang distribution, $P_{\rm Erlang}(\tilde{p}; D_S)$, given by,
\eq{
P_{\rm Erlang}(\tilde{p}; D_S) = \frac{D_S^{D_S}}{(D_S-1)!}e^{-D_S\tilde{p}}\tilde{p}^{D_S-1}\,.\label{eq:erlang}
}
For $|S|=0$ (equivalently $D_S=1$), the distribution above reduces to the Porter-Thomas distribution characteristic of pure-state designs.
More importantly, this Erlang distribution can experimentally signify the non-trivial moments of the projected ensemble and distinguish it from a trivial ensemble. For the latter (say, due to decoherence), $\varrho_A(z_B)\sim \mathbb{I}_{D_A}$ which would lead to ${\rm PoP}_{z_A}(\tilde{p})=\delta(\tilde{p}-1)$.

The emergence of the Erlang distribution for mixed-state deep thermalisation can be understood straightforwardly as the projected ensemble approaches a gHSe at late times. The gHSe is nothing but a partial trace of the Haar ensemble with the PoPs in the latter described by a Porter-Thomas distribution. The bitstring probabilities in the gHSe are then simply sums of $D_S$ Porter-Thomas distributed variables, which are distributed according to the Erlang distribution in Eq.~\ref{eq:erlang}.

\subsection{Determining second-order R{\'e}nyi entropies without randomised measurements}

As a practical application of mixed-state deep thermalisation, here we demonstrate a new method to estimate the second-order R{\'e}nyi entropy of an arbitrary many-body mixed state. 
The method relies on the fact that the bitstring PoPs at late times, which are experimentally accessible, encode information about the spectrum of the initial mixed state.

The method is analogous to the randomised measurement protocol~\cite{brydges2019randomized}, where local random unitaries sampled from a 2-design are applied on a mixed state, and the purity is estimated from the unitary-averaged cross-correlations of bitstring probabilities. 
By contrast, our method does not require any averaging over unitaries and is useful in practical settings where one is unable to sample sufficiently random ensembles of unitaries.

\begin{figure}[t]
\includegraphics[width=.5\linewidth]{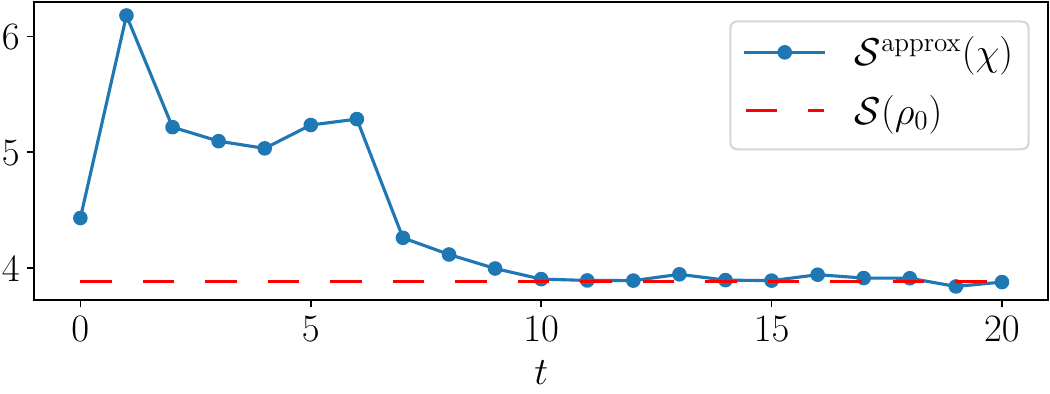}
    \caption{The proxy for the second R{\'e}nyi entropy from the bitstring probabilities (solid line) plotted against time for the kicked Ising model for $L=12$ and $|A|=3$. The dashed line denotes value of the second R{\'e}nyi entropy of the initial state $\rho_0$.}
    \label{fig:proxy}
\end{figure}

For instance, consider the evolution of a mixed state $\rho_0$ by the kicked Ising chain. To estimate the second R{\'e}nyi entropy of $\rho_0$, we define the correlator $\chi$ given by
\eq{\chi=\sum_{z_A,z_B}p(z_A,z_B)~p(z_A/z_B)\,,}
where $p(z_A,z_B)$ and $p(z_A/z_B)$ are the joint and conditional bitstring probabilities, respectively. The proxy for the second R{\'e}nyi entropy ${\cal S}_2(\rho_0)$ is defined by
\eq{{\cal S}^{\rm{approx}}_2(\chi)=-\ln\left[\frac{D_A-1}{1-\chi}-D_A\right]\,.
\label{eq:2re-proxy}}
Since the second moment of the mixed-state projected ensemble is approximately equal to that of ${\cal E}_{\rm {ref}}[\rho_0]$ after evolving the state for sufficiently long times, we have 
\eq{{\cal S}_2(\rho_0)\approx{\cal S}^{\rm{approx}}_2(\chi)\,,
\label{eq:proxy}}
which follows from Eq.~\ref{eq:eref-second-moment} of the main text. To verify Eq.~\ref{eq:proxy}, we consider the kicked Ising evolution of an arbitrary mixed state $\rho_0$ over $12$ qubits. In Fig.~\ref{fig:proxy}, we see that ${\cal S}^{\rm{approx}}_2(\chi)$ converges to ${\cal S}_2(\rho_0)$, demonstrating the utility of mixed state deep thermalisation in learning the finer features of quantum states.

\end{document}